\documentclass[aps,amssymb,showpacs,floatfix,twocolumn,prb,reprint,longbibliography]{revtex4-2}

\usepackage{subfiles} 
\usepackage[colorlinks=true, allcolors=blue]{hyperref}
\newcommand{\bra}[1]{\langle#1|} 
\newcommand{\ket}[1]{|#1\rangle} 

\newcommand{\vk}{\color{red}}

\usepackage{graphicx}
\usepackage{dcolumn}
\usepackage{bm}
\usepackage[caption=false]{subfig}
\usepackage{amsmath}
\usepackage[normalem]{ulem}

\usepackage{amssymb}
\usepackage{epsf}
\usepackage{xcolor}
\usepackage{tikz}

\def \bk{{\bf k}}
\def \be{\begin{equation}}
\def \ee{\end{equation}}
\newcommand{\ve}{\varepsilon}

\allowdisplaybreaks[1] 

\begin{document}

\title{
Strong photogalvanic effect in Weyl materials due to magnetic resonances}
\author{Vahid Hassanzade}
\affiliation{Department of Physics, Carnegie Mellon University, Pittsburgh, Pennsylvania 15213, USA}

\author{Vladyslav Kozii}
\affiliation{Department of Physics, Carnegie Mellon University, Pittsburgh, Pennsylvania 15213, USA}

\begin{abstract}
We study the photogalvanic effect in Weyl semimetals under a magnetic field, focusing on the shift current. Using the Kubo formalism for an ideal, clean Weyl node at zero temperature, we derive a general analytic expression valid for arbitrary light frequency, Fermi energy, and magnetic field strength. We identify a series of resonances that can be probed experimentally. To complement the microscopic analysis, we employ the semiclassical Boltzmann approach, which allows us to incorporate finite scattering phenomenologically. Unlike most previous studies using this method, we do not treat the magnetic field perturbatively; instead, we solve the Boltzmann equation for a Weyl node exactly within the limits of validity of the semiclassical theory. Our solution reproduces the low-frequency resonances and elucidates the role of finite scattering.
\end{abstract}
\maketitle

\section{Introduction}

Light-to-charge conversion underpins key optoelectronic technologies, including photovoltaics, photodetection, and sensing~\cite{Brongersma2015}. Topological semimetals, characterized by gapless spectra and nontrivial quantum geometry, have emerged as promising platforms for these applications~\cite{NagaosaSolarCells2017,photodetection2020}. Their nontrivial band geometry strongly enhances nonlinear optical responses, making them attractive candidates for nonlinear optoelectronic devices~\cite{Cook2017,Yang2018,Tan2016,Morimoto2016sciadv,Glazov2014,Okada2016}. These properties position topological semimetals as potential alternatives to conventional solar cells and photodetectors.

Large photocurrents in semimetals are commonly attributed to the bulk photogalvanic effect (PGE), a second-order rectified current generated under uniform illumination in noncentrosymmetric materials. Microscopically, the PGE originates from inversion-asymmetric changes in carrier position or velocity during optical excitation, leading to shift and injection currents, respectively~\cite{Sipe2000}. Among these, shift currents are particularly notable: they arise from a real-space displacement of the Wannier center following interband excitation, without relying on built-in electric fields or carrier density gradients~\cite{Young2012}. As a result, shift-current-based devices can circumvent the Shockley–Queisser limit that constrains conventional $p–n$ junction photovoltaics~\cite{Shockley1961}.

The intrinsic PGE is intimately connected to the geometry and topology of the electronic band structure, involving quantities such as Berry curvature, the quantum metric, and Christoffel symbols~\cite{Morimoto2016sciadv,AhnNagaosa2020}. Weyl semimetals provide a paradigmatic example of this connection: in systems lacking inversion and mirror symmetries, the circular photogalvanic effect (CPGE) is predicted to be nearly quantized and proportional to the monopole charge of the Weyl nodes~\cite{deJuan2017,Grushin2018,Reeseaba0509}. More broadly, the deep link between nonlinear optical responses and band geometry has established nonlinear optics as a powerful probe of electronic structure, motivating extensive theoretical and experimental efforts~\cite{Rostamis2018,Ji2019,Ma2017,Osterhoudt2019,Wu2017,Haibin2018,Zyuzin2017,Golub2017,Zhang2018,Chan2017}.

In this work, we present an analytical calculation of the shift current in a WSM in the presence of an external magnetic field. It is well established that magnetic fields can strongly enhance nonlinear optical responses in semimetals, highlighting magneto-optics as a promising avenue for technological applications~\cite{GolubIvchenko2018,Yao_2013,Belyanin2012,Ornigotti_2023,BednikKozii2024}. In recent work, one of us showed that both the injection current and second-harmonic generation in WSMs are significantly amplified by a magnetic field, giving rise to a series of magnetic resonances~\cite{BednikKozii2024}. Here, we complete the analysis of the second-order conductivity by evaluating the intrinsic contribution to the shift current.


Using an idealized model of a linear, isotropic three-dimensional band crossing, we employ the Kubo formula to study a clean, noninteracting system. We derive general zero-temperature expressions for the shift current components valid for arbitrary incident light frequency, chemical potential, and magnetic field, as long as all relevant energy scales remain below the cutoff beyond which the linear dispersion approximation breaks down. Our results show that even relatively weak magnetic fields can dramatically enhance the nonlinear optical response, leading to divergent shift currents associated with magnetic resonances.


We complement our analysis with a semiclassical treatment based on the Boltzmann equation, which elucidates the role of single-particle scattering. In contrast to most previous studies, we treat the magnetic field nonperturbatively, alongside the light frequency and scattering rate. We solve the Boltzmann equation exactly to second order in the electric field within the relaxation-time approximation while neglecting interband transitions. Neglecting interband transitions is expected to be justified when the Fermi energy is the largest energy scale below the cutoff. We find that the results for second-harmonic generation are in good quantitative agreement with those obtained from the Kubo-formula approach. In contrast, the agreement for the shift current is at best qualitative in certain limits and absent in others. The origin of this discrepancy remains an interesting open question that warrants further investigation.

The rest of the paper is organized as follows. In Sec.~\ref{Model}, we introduce definitions and our model followed by the outline of the computation steps. In Sec.~\ref{photocurrent}, we derive the analytical expressions for the shift current components from the Kubo formula in the clean limit, identify the divergences, and discuss different limiting cases. We present our semiclassical results in Sec.~\ref{semiclassics}. We make the final remarks in Sec.~\ref{Conclusion}. We discuss various details of our calculations in numerous appendices.

\begin{figure}
    \centering
    \includegraphics[width=0.85\linewidth]{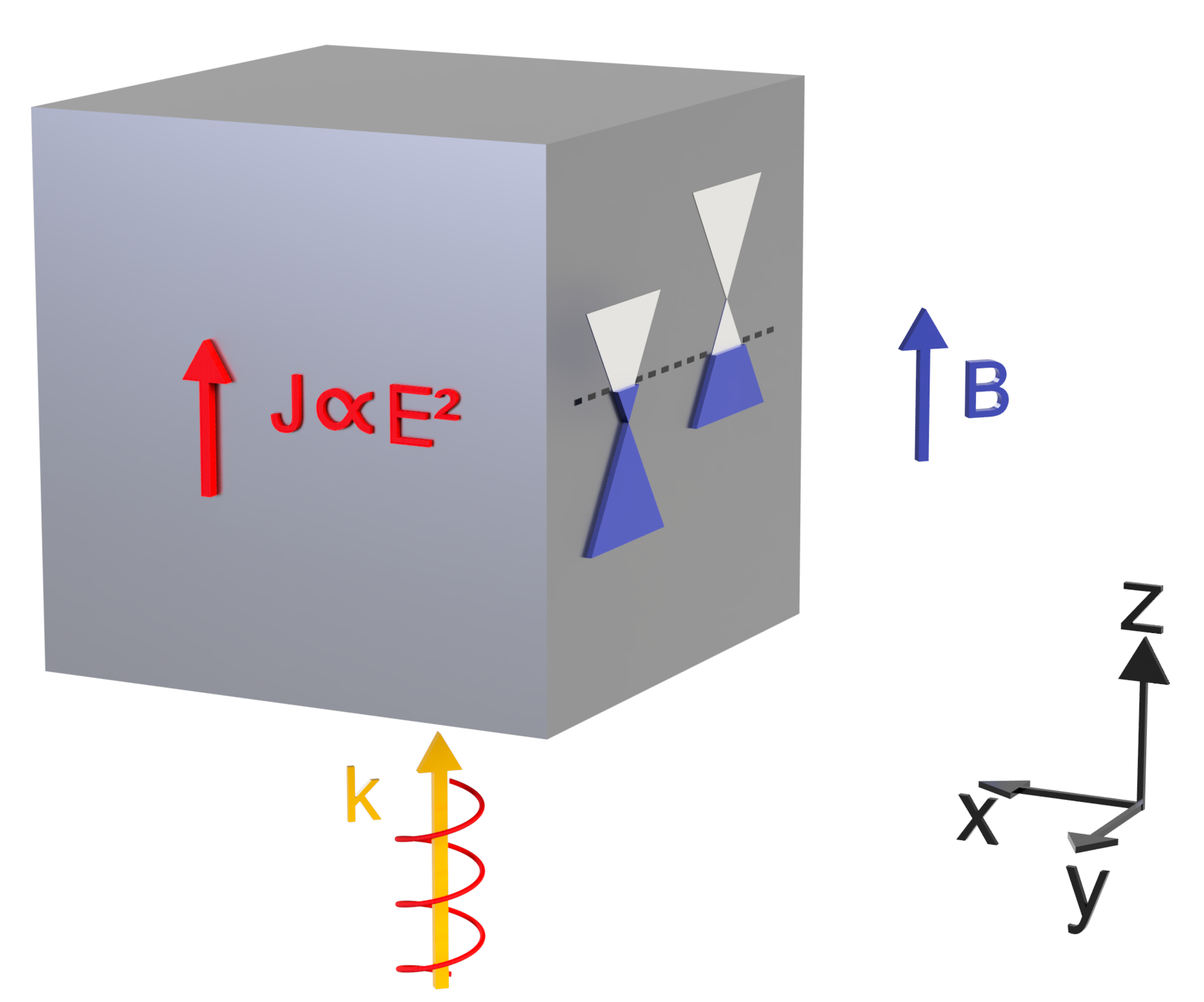}
    \caption{Schematic picture of the setup presented in the paper. A Weyl semimetal with nodes at different energies is considered under a magnetic field applied along the $z$ direction. The image depicts the photogalvanic effect (PGE), with incident light propagating along the $z$-axis and possessing electric field components in the $x$–$y$ plane impinging on the sample. As a result the photocurrent $J$ flows in the $z$ direction.}
    \label{fig:setup}
\end{figure}

\section{Model \label{Model}}

Our calculation closely follows that of Ref.~\cite{BednikKozii2024}. In particular, we start with a single isotropic Weyl node with linear dispersion. Assuming that the magnetic field B($>0$) is along the $z$-axis as shown in Fig.~\ref{fig:setup}, we use the Landau gauge $\mathbf{A}=(-yB,0,0)$ and write the Hamiltonian as

\begin{equation}
    \hat{H} = \eta v_f \hbar \bigg[\big( k_x - eBy/\hbar \big)\sigma_x -i\partial_y\sigma_y +k_z\sigma_z \bigg].
    \label{hamiltonian}
\end{equation}
Here, $\eta=\pm 1$ determines the chirality of the node, $v_f$ is the Fermi velocity near the Weyl node, $-e$ is the electron charge, and $\boldsymbol \sigma$ is the vector of Pauli matrices. Translational invariance of the Hamiltonian in the $x-$ and $z-$directions implies that the wavevector components $k_x$ and $k_z$ remain good quantum numbers. 

We emphasize that our analysis accounts solely for the orbital effects of the magnetic field. The Zeeman coupling to the electron spin is neglected, which is justified when the effective $g$-factor is sufficiently small. The inclusion of Zeeman effects and their interplay with orbital dynamics may lead to qualitatively new phenomena~\cite{Dzeroteal2025,Dzero-Rashba25}, which we leave for future work.

Weyl nodes in WSMs always occur in pairs with opposite chiralities. As we show below, the shift-current contribution from an individual node is proportional to its chirality. Consequently, a nonzero net current requires an energy separation between nodes of opposite chirality. This, in turn, demands that the WSM breaks not only inversion symmetry — as required for any second-order response — but also all mirror symmetries~\cite{deJuan2017}, an assumption we adopt throughout this work.

The generic second-order current is expressed through  the nonlinear conductivity tensor $\sigma^{\alpha\beta\gamma}(\omega_1,\omega_2)$ as 
\begin{equation}
    j^\gamma(\Omega) = \sigma^{\alpha\beta\gamma}(\omega_1,\omega_2) E^\alpha(\omega_1)E^\beta(\omega_2),
\end{equation}
where $E^\alpha(\omega_{1,2})$ are the electric field components with frequency $\omega_{1,2}$, $\Omega \equiv \omega_1+\omega_2$, and we imply the summation over the repeated indices $\alpha, \beta = x,y,z$. We follow the steps of Refs.~\cite{Avdoshkin2020,BednikKozii2024} and use the Matsubara technique to perform our analytical calculation. The second-order conductivity tensor can then be expressed through the three-point correlation function $\chi^{\alpha\beta\gamma}$ as 
\begin{equation}
    \sigma^{\alpha\beta\gamma}(i\omega_1,i\omega_2) = \frac{\chi^{\alpha\beta\gamma}(i\omega_1,i\omega_2)+\chi^{\beta\alpha\gamma}(i\omega_2,i\omega_1)}{\omega_1\omega_2},
    \label{eq-sigma_xi}
\end{equation}
where in the limit of zero temperature
\begin{align}
    &\chi^{\alpha\beta\gamma}(i\omega_1,i\omega_2) = \label{xi}\\ \nonumber
    &\frac{1}{V}\int \frac{d\varepsilon}{2\pi}\text{Tr} \left[\hat{j}^\alpha G(i\varepsilon-i\omega_1)\hat{j}^\beta G(i\varepsilon-i\Omega)\hat{j}^\gamma G(i\varepsilon)\right].
\end{align}
Here, $V$ denotes the system’s volume, and the trace indicates the integration over intermediate coordinates as well as summation over (pseudo-)spin indices. The exact Green’s function in the presence of a uniform external magnetic field is given by

\begin{equation}
    G(i\varepsilon,\mathbf{r}_1,\mathbf{r}_2) = \sum_{n,k_x,k_z}\frac{\ket{\Psi_{n,k_x,k_z}(\mathbf{r}_1)}\bra{\Psi_{n,k_x,k_z}(\mathbf{r}_2)}}{i\varepsilon-E_n(k_z)+\mu}.
\end{equation}
Here, $\ket{\Psi_{n,k_x,k_z}(\mathbf{r})}$ denotes the eigenstates of Hamiltonian~\eqref{hamiltonian}, labeled by the quantum numbers $n, k_x,$ and $k_z$. Its coordinate-space functional representation can be found, e.g., in Appendix~A of Ref.~\cite{BednikKozii2024}. The eigenenergies $E_n(k_z)$ do not depend on $k_x$ and are given by
\begin{equation}
    E_n(k_z) = \begin{cases}
        \text{sgn}(n)\eta\hbar\sqrt{v_f^2k_z^2+\omega_B^2|n|}, & n \neq 0,\\
        -\eta \hbar v_f k_z,  & n = 0, 
        \end{cases}
    \label{eq:energy spectrum}
\end{equation}
where $\omega_B^2=2eBv_f^2/\hbar$ is the square of the characteristic magnetic frequency and $E_0(k_z)$ is the single chiral band. The chemical potential $\mu$ is defined relative to the node’s position.

The expression for the current operator is

\begin{equation}
    \hat{j}^\alpha = -\frac{e}{\hbar}\frac{\delta H_\mathbf{k}}{\delta k^\alpha} = -\eta e v_f \sigma_\alpha,
\end{equation}
where $H_\mathbf{k} = \eta \hbar v_f {\bf k} \cdot {\boldsymbol \sigma}$ is the band Hamiltonian. Higher-order derivatives vanish near a Weyl node, $\delta^2 H_\bk/\delta k^\alpha \delta k^\beta =0$, so they do not contribute to current.

Carrying out integration over $\varepsilon$ in Eq.~\eqref{xi} gives

\begin{widetext}
\begin{equation}
    \chi^{\alpha\beta\gamma}(i\omega_1,i\omega_2) = (ev_f)^3\frac{\eta e B}{2\pi \hbar}\int_{-\infty}^{+\infty} \frac{dk_z}{2\pi}\sum_{n_1,n_2,n_3}Z^{\alpha\beta\gamma}_{n_1,n_2,n_3} \frac{1}{i\hbar\omega_1+\varepsilon_{n_1}-\varepsilon_{n_3}}\bigg[\frac{\Theta(\varepsilon_{n_2})-\Theta(\varepsilon_{n_1})}{i\hbar\omega_2+\varepsilon_{n_2}-\varepsilon_{n_1}} - \frac{\Theta(\varepsilon_{n_2})-\Theta(\varepsilon_{n_3})}{i\hbar\Omega+\varepsilon_{n_2}-\varepsilon_{n_3}}\bigg],
    \label{xi2}
\end{equation}
\end{widetext}
where we defined $\varepsilon_{n_i} = E_{n_i}(k_z)-\mu$, $\Theta(\varepsilon)$ is the Heaviside step function, and 
\begin{equation}
    Z^{\alpha\beta\gamma}_{n_1,n_2,n_3} = \bra{\Psi_{n_3}}\sigma_\alpha\ket{\Psi_{n_1}} \bra{\Psi_{n_1}}\sigma_\beta\ket{\Psi_{n_2}}\bra{\Psi_{n_2}}\sigma_\gamma\ket{\Psi_{n_3}}.
\end{equation}
The explicit expressions for these matrix elements can again be found in Appendices of Ref.~\cite{BednikKozii2024}.

To obtain physical conductivity, we will perform analytic continuation $i\omega_{1,2} \to \omega_{1,2}+i0$ from the upper complex half-plane to the real frequency axis. Equation~\eqref{xi2} provides the starting point for the Kubo formula calculation of the shift current in the next section.

In the zero-field limit, $B=0$, Eq.~\eqref{xi2} for an ideal Weyl node predicts that only the chiral photocurrent associated with the circular photogalvanic effect (CPGE) is nonzero, while the linear photogalvanic effect (LPGE) and second-harmonic generation vanish~\cite{Avdoshkin2020,BednikKozii2024}. This result is explicitly derived in Appendix~\ref{AppB}. This behavior reflects the high emergent symmetry of the ideal Weyl-node model. A finite magnetic field partially lifts this symmetry, leading to a nonzero LPGE response.




\section{Shift current}
\label{photocurrent}

The purpose of this section is to compute the photocurrent, defined as the dc component of the second-order current generated by monochromatic light of frequency~$\omega$. To extract the corresponding conductivity tensor, we analytically continue Eqs.~\eqref{eq-sigma_xi} and~\eqref{xi2} from the upper half-plane of complex frequencies according to~\footnote{The analytic continuation symmetric with respect to $\Omega$, $i\omega_1\to\omega+\Omega/2+i0$ and $i\omega_2\to-\omega+\Omega/2+i0$, significantly simplifies the calculation and automatically eliminates unphysical contributions to the photocurrent.} 
\begin{align} 
\lim_{\Omega\to 0} &\sigma^{\alpha \beta \gamma}\left(\omega + \frac{\Omega}2 + i 0, -\omega + \frac{\Omega}2 + i 0 \right)  \nonumber \\ =&\frac{i}{\Omega} \beta^{\alpha \beta \gamma}(\omega) + \sigma^{\alpha \beta \gamma}(\omega).
\end{align}
The term proportional to $1/\Omega$ corresponds to a current component that grows linearly in time in a clean, noninteracting system and describes the injection current~\cite{Sipe2000}. In realistic systems, this growth is cut off by single-particle scattering, leading to saturation at long times~\cite{deJuan2017,Levchenko2017}. This contribution to conductivity has been evaluated previously in Refs.~\cite{GolubIvchenko2018,BednikKozii2024}. The second term, which is independent of $\Omega$, describes the shift current~\cite{Sipe2000} and constitutes the primary focus of the present section. While the decomposition into injection and shift currents — traditionally associated with changes in carrier velocity and position — becomes ambiguous in the presence of a magnetic field, we retain this terminology for $\beta^{\alpha \beta \gamma}$ and $\sigma^{\alpha \beta \gamma}$ for clarity and convenience.


Within our model, the shift current flows exclusively along the magnetic-field direction, i.e., the $z-$axis. Specifically, the only nonvanishing components of the conductivity tensor are $\sigma^{xxz}(\omega)=\sigma^{yyz}(\omega)$  and $\sigma^{xyz}(\omega)=-\sigma^{yxz}(\omega)$. To evaluate these components, we analytically continue Eqs.~\eqref{xi2} and~\eqref{eq-sigma_xi} in Appendix~\ref{AppA}, expand Eq.~\eqref{eq-sigma_xi} in powers of $\Omega$, and perform the integration over $k_z$ in Appendix~\ref{AppC}. 
\subsection{$\sigma^{xxz}(\omega)$ component of the shift current}
The analytic expression for the LPGE component of the shift current conductivity is given by 
\begin{align} \label{xxz-sigma}
   &\sigma^{xxz}(\omega)=\sigma^{yyz}(\omega) \\ &= \frac{\eta e^3 }{8\pi^2 \hbar^2} \frac{\omega_B^2 \text{sgn}(\mu)}{\omega^2}
    \sum_{n=0}^\infty
    \left(\frac{k_n}{k_n^2+p_n^2}-\frac{k_{n+1}}{k_{n+1}^2+p_n^2}\right), \nonumber
\end{align} 
where we have defined

\begin{equation}
    k_n \equiv \text{Re}\sqrt{\frac{\mu^2}{\hbar^2}-n\omega_B^2}, \hspace{0.5cm}
    p_n^2 \equiv n\omega_B^2- \frac{(\omega^2-\omega_B^2)^2}{4\omega^2}.
    \label{definitions}
\end{equation}
Quantity $\hbar k_n/v_f$ has meaning of the momentum at which the chemical potential $\mu$ intersects the $n$-th Landau level.

\begin{figure}
\centering
\includegraphics[width=1\linewidth]{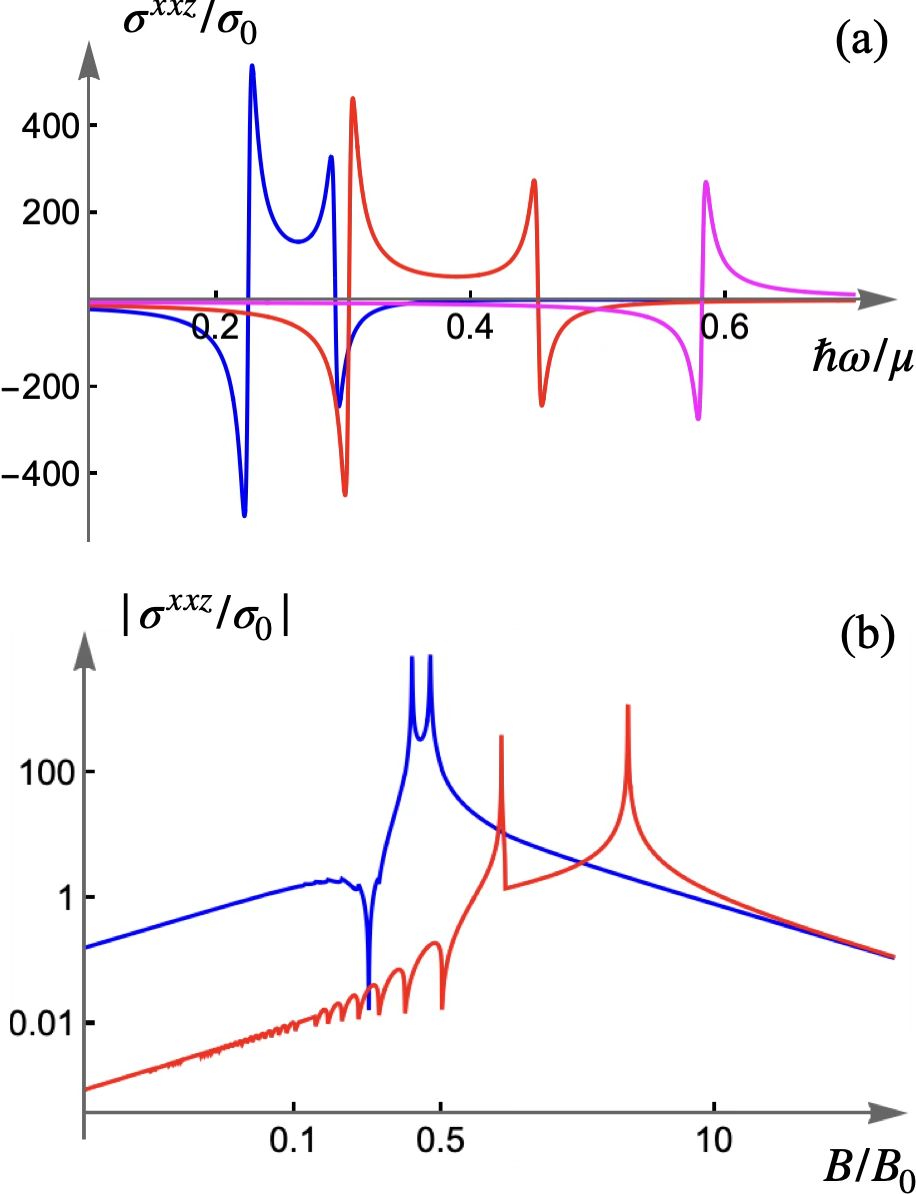}
\caption{\label{fig-xxz}  The second order dc shift photocurrent $\sigma^{xxz}$ as a function of a) incident light frequency $\omega$, and b) magnetic field $B$ calculated from Eq. \eqref{xxz-sigma} at fixed chemical potential $\mu$. Here, $\sigma_0 \equiv \eta e^3/16\pi^2\hbar\mu$ and $B_0 \equiv \mu^2/2e\hbar v_f^2$. In panel (a), different curves correspond to different values of the magnetic field: blue ($B/B_0 = 0.5$), red ($B/B_0 = 0.7$), and magenta ($B/B_0 = 1.5$), where $B/B_0 = \hbar^2 \omega_B^2 / \mu^2$. In panel (b), a double-logarithmic scale is employed to highlight distinct dynamical regimes. The curves represent different values of the incident light frequency: blue ($\hbar\omega/\mu = 0.2$) and red ($\hbar\omega/\mu = 1.2$). A small but finite imaginary part of frequency is kept to regularize the divergences in the presented graphs. Quantum oscillations are seen at intermediate fields.}
\end{figure}

Figure~\ref{fig-xxz} shows the corresponding conductivity component as a function of the incident light frequency~$\omega$ and magnetic field~$B$. Strictly speaking, the field dependence should include the variation of the chemical potential with magnetic field, $\mu \to \mu(B)$, which follows from fixing the total particle number. This dependence is, however, sensitive to the relative positions of the Weyl nodes and the Fermi energy, since any physical realization necessarily involves multiple nodes of opposite chirality located at different energies. We defer a detailed analysis of this effect to future work on specific materials with realistic band structures, and for simplicity treat $\mu$ as an externally fixed parameter here.

Equation~\eqref{xxz-sigma} indicates that the $\sigma^{xxz}(\omega)$ component exhibits power-law divergencies at resonant frequencies 
\begin{align}  \label{freq_w}
 &\hbar\left|\omega_1^{\pm}\right| = \sqrt{\mu^2+\hbar^2\omega_B^2} \pm \mu, \\ 
 &\hbar\left|\omega_2^{\pm}\right| = |\mu| \pm \sqrt{\mu^2-\hbar^2\omega_B^2}, \qquad (\text{if}\,\,\, \hbar\omega_B <|\mu|). \nonumber
\end{align}
These divergences are cut off by the nonzero scattering rate arising from disorder or interactions, as we demonstrate in the next section using the Boltzmann equation formalism. The relation between these frequencies and resonant transitions between Landau levels is discussed, for example, in Ref.~\cite{BednikKozii2024}.

In the limit of small magnetic fields, $\hbar\omega_B\ll\min\{\hbar\omega,\mu, \sqrt{|\hbar^2\omega^2-4\mu^2|}\}$, we find that 
\begin{equation} \label{Eq:xxz_smallB}
    \sigma^{xxz}(\omega) \approx \frac{\eta e^3 \omega_B^2 \mu (4\mu^2-3\hbar^2\omega^2)}{6\pi^2\hbar\omega^2 \big( 4\mu^2-\hbar^2\omega^2\big)^2
    },
\end{equation}
and at small frequencies, $\hbar\omega\ll \min\{\hbar\omega_B,\hbar^2\omega_B^2/\mu\}$, we obtain
\begin{equation}\label{Eq:xxz_loww}
    \sigma^{xxz}(\omega) \approx -\frac{\eta e^3 \mu}{2\pi^2 \hbar^3 \omega_B^2}.
\end{equation}
We also clearly see quantum oscillations at intermediate fields in Fig.~\ref{fig-xxz}. 


\subsection{$\sigma^{xyz}(\omega)$ component of the shift current}
\begin{figure}
\includegraphics[width=1\linewidth]{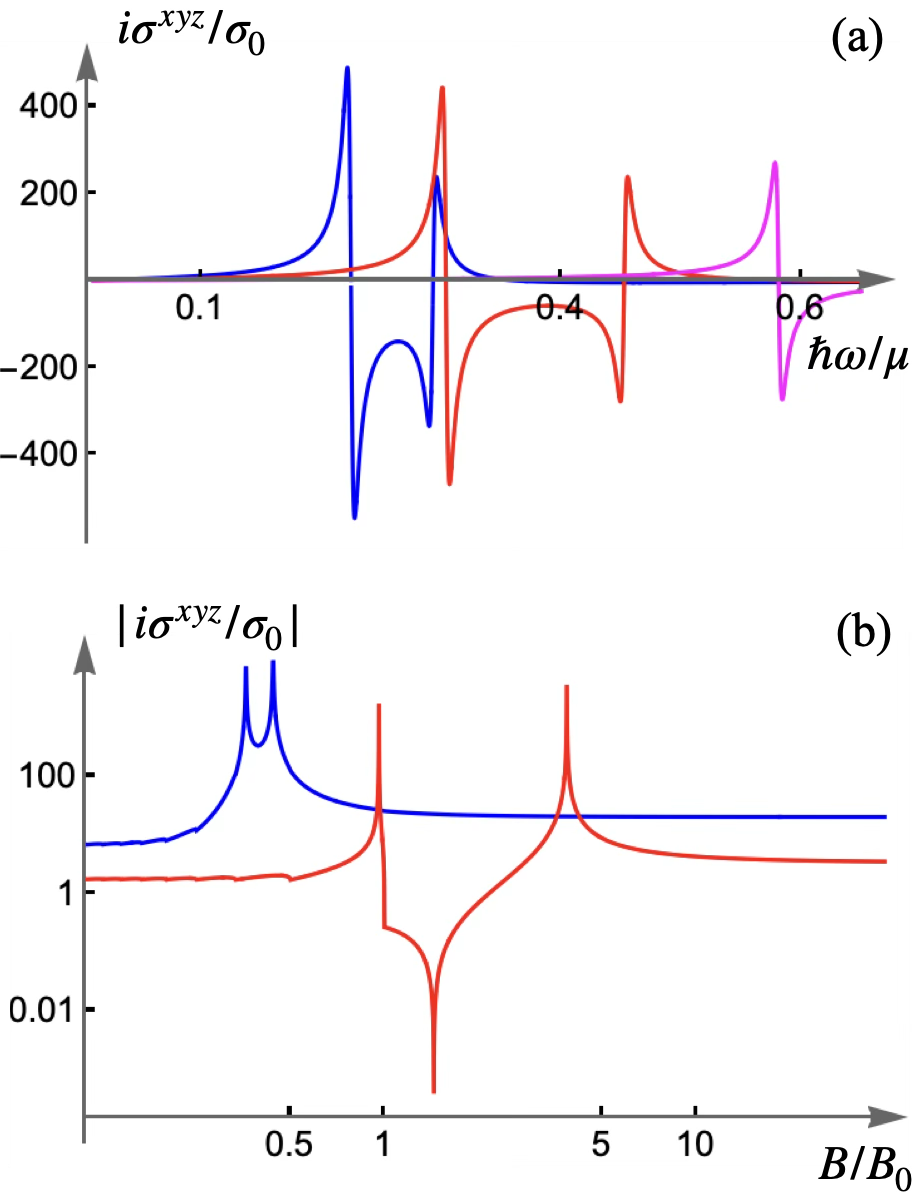}
\caption{\label{fig-xyz} Same graphs as in Fig.~\ref{fig-xxz} but for the $\sigma^{xyz}$ component of the shift photocurrent, given by Eq.~\eqref{xyz-sigma}.}
\end{figure}

The CPGE component of the shift current conductivity equals
\begin{align}
    &\sigma^{xyz}(\omega) =-\sigma^{yxz}(\omega) \label{xyz-sigma}
\\
    &=-\frac{i\eta e^3 }{16\pi^2\hbar}\frac{\omega_B^2 }{\omega^3|\mu|}
    \sum_{n=0}^\infty\left[
    \frac{k_n(\omega^2-\omega_B^2)}{k_n^2+p_n^2}+\frac{k_{n+1}(\omega^2+\omega_B^2)}{k_{n+1}^2+p_n^2}\right],
     \nonumber
\end{align}
where $k_n$ and $p_n$ are defined in Eq.~(\ref{definitions}). Figure~\ref{fig-xyz} shows conductivity component $\sigma^{xyz}(\omega)$ as a function of the incident light frequency~$\omega$ and magnetic field strength~$B$. The divergences occur at the same resonant frequencies $\omega_1^\pm$ and $\omega_2^\pm$ given by Eq.~\eqref{freq_w}.

In the limit of weak magnetic fields, $\hbar\omega_B\ll\min\{\hbar\omega,\mu, \sqrt{|\hbar^2\omega^2-4\mu^2|}\}$, the conductivity saturates at a finite value,
\begin{equation} \label{Eq:xyz_smallB}
    \sigma^{xyz}(\omega) \approx \frac{i\eta e^3\mu^2}{3\pi^2\hbar^2\omega(\hbar^2\omega^2 - 4\mu^2)}.
\end{equation}
At first sight, this result appears inconsistent with the exact photoconductivity obtained in the absence of a magnetic field. As shown in Appendix~\ref{AppB}, based on the results of Refs.~\cite{Avdoshkin2020,BednikKozii2024}, the corresponding conductivity component at $B=0$ is
\be  \label{Eq:xyzB=0}
\sigma^{xyz}(\omega,B=0) = \frac{i\eta e^3 \omega}{12 \pi^2 (\hbar^2 \omega^2 - 4\mu^2)}.
\ee 

The apparent discrepancy originates from the noncommutativity of the limits $B\to0$ and $\Omega\to0$. Equation~\eqref{Eq:xyz_smallB} is obtained by first taking the dc limit, $\Omega\to0$, and only then sending $B\to0$, corresponding to the clean, noninteracting system in the presence of an infinitesimal magnetic field. In contrast, Eq.~\eqref{Eq:xyzB=0} is obtained by setting $B=0$ from the outset. Thus, the two expressions correspond to different orders of limits, which do not commute. Importantly, the difference between the two results is
\be  \label{Eq:sigma-sigma}
\sigma^{xyz}(\omega, B=0) - \sigma^{xyz}(\omega, B\to 0)  = \frac{i \eta e^3}{12 \pi^2 \hbar^2 \omega},
\ee 
and does not depend on the chemical potential. Consequently, it cancels upon summing over Weyl nodes of opposite chirality. We therefore conclude that, for a physically relevant Weyl semimetal containing equal numbers of nodes with opposite chiralities, the value of $\sigma^{xyz}(\omega)$ is ultimately insensitive to the order of limits.

We also note that Eq.~\eqref{Eq:xyzB=0} applies to all conductivity components related to $\sigma^{xyz}(\omega)$ by cyclic permutations of the coordinate indices. In contrast, Eqs.~\eqref{xyz-sigma}-\eqref{Eq:xyz_smallB} describe only the components with the current flowing along the $z$ direction, while all other CPGE components vanish. As a result, the cancellation discussed above is specific to $\sigma^{xyz}(\omega)$. For other components, such as $\sigma^{yzx}(\omega)$ and $\sigma^{zxy}(\omega)$, the order of limits generally remains important. This behavior is fully analogous to that found for the injection current in Ref.~\cite{BednikKozii2024}.

Finally, we find for small frequencies $\hbar\omega\ll \min\{\hbar\omega_B,\hbar^2\omega_B^2/\mu\}$ 
\begin{equation}
    \sigma^{xyz}(\omega) \approx -\frac{i\eta e^3}{4\pi^2\hbar^2 \omega}\bigg[1+\frac{\omega^2}{\omega_B^2}\bigg( \frac{4\mu^2}{\hbar^2\omega_B^2}+1+2\sum_{n=1}^\infty\frac{\hbar k_n}{|\mu|} \bigg) \bigg] .
\end{equation}
If we additionally assume that $\hbar \omega_B \ll \mu$, this expression further simplifies to 
\begin{equation} \label{Eq:xyz_loww_lowB}
    \sigma^{xyz}(\omega) \approx -\frac{i\eta e^3}{4\pi^2\hbar^2 \omega}\bigg(1+\frac{16\omega^2\mu^2}{3\hbar^2\omega_B^4} \bigg) 
\end{equation}
The first (leading) term in this low-frequency expansion also cancels out after summation over the nodes with opposite chiralities, similarly to Eq.~\eqref{Eq:sigma-sigma}.  





\section{Semiclassical Approach}
\label{semiclassics}
We next use the semiclassical Boltzmann equation to calculate the second-order response, including both the PGE current and second-harmonic generation (SHG) in Weyl semimetals. As is standard in the Boltzmann approach, we focus on the band containing the Fermi energy and neglect interband and internodal processes. This approximation is generally justified in the regime $\max\{\hbar\omega_B,\hbar\omega\}\ll|\mu|$ which we assume from now on, along with the Mott-Ioffe-Regel criterion $\mu \tau \gg \hbar$. As we show below, however, only the SHG response obtained within the semiclassical approach is in good agreement with the microscopic Kubo-formula calculation in the clean limit. By contrast, the results for the PGE current agree only qualitatively in some limits and differ substantially in others.


We extend the previous semiclassical analysis of Ref.~\cite{BednikKozii2024} to treat the magnetic field non-perturbatively, while still assuming $\hbar\omega_B\ll|\mu|$. By solving the Boltzmann equation {\it exactly} to second order in the electric field, we can describe the low-frequency regime and elucidate the role of a finite single-particle scattering rate near  low-frequency resonances $\omega \approx \hbar\omega_B^2/2\mu,\, \hbar\omega_B^2/4\mu$.

We start with the semiclassical equations of motion that for an electron in a solid have the form \cite{Xiaoreview2010}
\begin{align}
    \hbar\dot{\mathbf{r}} &= \mathbf{\nabla_k}\varepsilon_\mathbf{k} - \hbar\dot{\mathbf{k}}\times\boldsymbol{\Omega}_\mathbf{k},
    \nonumber\\
    \hbar\dot{\mathbf{k}} &= -e\mathbf{E} - e \dot{\mathbf{r}}\times\mathbf{B},
    \label{motion_1}
\end{align}
where
\begin{align}
    \mathbf{\Omega_k} = i\bra{\nabla_\mathbf{k} u_\mathbf{k}}\times \ket{\nabla_\mathbf{k} u_\mathbf{k}}
\end{align}
is the Berry curvature and $\ket{u_\mathbf{k}}$ is the periodic part of the Bloch wave function. The quasiparticle dispersion relation is modified in the presence of magnetic field, $\varepsilon_\mathbf{k}=\varepsilon_\mathbf{k}^0-\mathbf{m_k}\cdot\mathbf{B}$, where $\varepsilon_\mathbf{k}^0$ is the bare band energy at $\mathbf{B}=0$ satisfying the Schr\"{o}dinger equation $H_\mathbf{k} \ket{u_\mathbf{k}} = \varepsilon_\mathbf{k}^0\ket{u_\mathbf{k}}$ and the orbital magnetic moment $\mathbf{m_k}$ is defined as
\begin{align}
    \mathbf{m_k} = -\frac{i e}{2\hbar}\bra{\nabla_\mathbf{k} u_\mathbf{k}}\times (H_\mathbf{k}-\varepsilon_{\mathbf{k}}^0)\ket{\nabla_\mathbf{k} u_\mathbf{k}}.
\end{align}

Equations~\eqref{motion_1} can be easily solved~\cite{Morimoto2016}:
\begin{align}
    \dot{\mathbf{r}} &= \frac{1}{\hbar D_{\mathbf{k}}} \left\{ \nabla_{\mathbf{k}} \varepsilon_{\mathbf{k}} + e \mathbf{E} \times \bm{\Omega}_{\mathbf{k}} + \frac{e}{\hbar} \mathbf{B} (\bm{\Omega}_{\mathbf{k}} \cdot \nabla_{\mathbf{k}} \varepsilon_{\mathbf{k}}) \right\}, 
    \nonumber\\
    \dot{\mathbf{k}} &= \frac{1}{\hbar D_{\mathbf{k}}} \left\{ -e \mathbf{E} - \frac{e}{\hbar} \nabla_{\mathbf{k}} \varepsilon_{\mathbf{k}} \times \mathbf{B} - \frac{e^2}{\hbar} (\mathbf{E} \cdot \mathbf{B}) \bm{\Omega}_{\mathbf{k}} \right\},
\end{align}
where $D_\mathbf{k} = 1+ (e/\hbar)(\mathbf{B} \cdot \mathbf{\Omega_k})$ is the phase-space volume correction due to a finite Berry curvature \cite{XiaoDOS2005}.

The time evolution of the distribution function of electrons in the presence of a uniform external electric field is governed by the Boltzmann equation, which in $\tau$-approximations takes form
\begin{align}
    \frac{\partial f}{\partial t} + \dot{\mathbf{k}} \cdot \nabla_\mathbf{k}f = -\frac{f-f_0}{\tau}.
    \label{boltzman}
\end{align}
Here we assume that the scattering rate $\tau$ is constant and $f_0 = \Theta(\mu-\varepsilon_\mathbf{k})$ is the zero-temperature Fermi-Dirac distribution function. The electric current can then be written as 
\begin{equation}
    \mathbf{j} = -e \int \frac{d^3k}{(2\pi)^3} D_{\mathbf{k}} \dot{\mathbf{r}} f. 
    \label{current-1}
\end{equation}
In case of a monochromatic light, $\mathbf{E}=\mathbf{E}_\omega e^{-i\omega t}+\mathbf{E}^*_\omega e^{i\omega t}$, the distribution function can be expanded to the second order in $E_\omega$ as
\begin{equation}
    f(t) \approx f_0 + \delta f_0 + (f_1 e^{-i\omega t} + f_1^* e^{i\omega t}) + (f_2 e^{-2i\omega t} + f_2^* e^{2i\omega t}),
\end{equation}
where $f_1 \propto E_\omega$ and  $\delta f_0, f_2 \propto E_\omega^2$, while we neglect terms of order $O(E_\omega^3)$. Separating various harmonics in Eq.~\eqref{boltzman}, we get the set of three differential equations:
\begin{widetext}
\begin{align}
    D_{\mathbf{k}} \left( i\omega - \frac{1}{\tau} \right) f_1 &= -\frac{e}{\hbar} \left[ \mathbf{E}_\omega + \frac{e}{\hbar} (\mathbf{E}_\omega \cdot \mathbf{B}) \boldsymbol{\Omega}_\mathbf{k} \right] \cdot \nabla_{\mathbf{k}} f_0 
    - \frac{e}{\hbar^2} \nabla_{\mathbf{k}} f_1 \cdot \nabla_{\mathbf{k}} \varepsilon_{\mathbf{k}} \times \mathbf{B}, \nonumber
    \\
    D_{\mathbf{k}}\left(2i\omega - \frac{1}{\tau}\right)f_2 &= -\frac{e}{\hbar} \left[\mathbf{E}_\omega + \frac{e}{\hbar}(\mathbf{E}_\omega \cdot \mathbf{B})\mathbf{\Omega_k} \right] \cdot \nabla_{\mathbf{k}} f_1
    - \frac{e}{\hbar^2} \nabla_{\mathbf{k}} f_2 \cdot \nabla_{\mathbf{k}} \varepsilon_{\mathbf{k}} \times \mathbf{B},
    \label{Diff_f2}
    \\
    - D_{\mathbf{k}} \frac{1}{\tau} \delta f_0 &= -\frac{e}{\hbar} \left[ \mathbf{E}_\omega^* + \frac{e}{\hbar} (\mathbf{E}_\omega^* \cdot \mathbf{B}) \boldsymbol{\Omega}_\mathbf{k} \right] \cdot \nabla_{\mathbf{k}} f_1 
    - \frac{e}{2\hbar^2} \nabla_{\mathbf{k}} \delta f_0 \cdot \nabla_{\mathbf{k}} \varepsilon_{\mathbf{k}} \times \mathbf{B} + \text{c.c.} \nonumber
\end{align}
\end{widetext}

Unlike Ref.~\cite{BednikKozii2024} and other related works that solved these differential equations perturbatively in powers of $B$, we find their {\it exact} solution for Weyl fermions. In this case, the bare band dispersion, the Berry curvature, and the orbital magnetic moment are given by 
\be
 \varepsilon_\mathbf{k}^0=\zeta v_f \hbar k, \quad \mathbf{\Omega_k}=-\frac{\zeta\eta}{2k^3}\mathbf{k}, \quad \mathbf{m_k}=-\eta\frac{ev_f}{2k^2}\mathbf{k},
 \ee
where $\zeta = +1/-1$ corresponds to the conduction/valence band and $\eta = \pm 1$ is the chirality of the node. The high symmetry of an ideal Weyl node allows us to transform these complicated partial differential equations into much simpler ordinary differential equations with respect to the azimuthal angle $\phi$. The resulting expressions for the solution and  current harmonics are still rather lengthy, and we present them fully in Appendix~\ref{appendix-D}. Here we discuss various limiting cases for both photocurrent and second-harmonic generation.

\subsection{Photocurrent}
Using the exact solution of Eqs.~\eqref{Diff_f2} for $f_1$ and $\delta f_0$ and inserting it in Eq.~\eqref{current-1}, we find the conductivity components of PGE that can be further evaluated in various limits. We start with the limit of the vanishing magnetic field, $B \to 0$, assuming again that it is pointing along the $z-$direction. Expanding to the third order in $B$ (sixth order in $\omega_B$) and keeping the leading contributions in each component, we find


\begin{align}   \label{semi-mu-limit}
   \sigma^{xxz}_\text{dc} &=\sigma^{yyz}_\text{dc} \approx -\frac{\eta e^3 \omega_B^2 \tau^2}{6\pi^2\hbar\mu}\frac{1}{(1+\omega^2\tau^2)^2},
    \\
    \nonumber\sigma^{xzx}_\text{dc} &=\sigma^{yzy}_\text{dc}=(\sigma^{zxx}_\text{dc})^*=(\sigma^{zyy}_\text{dc})^* =
    \\
   \nonumber &\approx \frac{\eta e^3 \omega_B^2 \tau^2}{24\pi^2\hbar\mu}\frac{2+i\omega\tau(1-\omega^2\tau^2)}{(1+\omega^2\tau^2)^2}, \\  \nonumber
    \sigma^{yzx}_\text{dc} &=(\sigma^{zyx}_\text{dc})^*=-\sigma^{xzy}_\text{dc}=-(\sigma^{zxy}_\text{dc})^* = \\ \nonumber &\approx \frac{\eta e^3 \omega_B^4 \omega \tau^4}{48\pi^2\mu^2}\frac{i(\omega^4 \tau^4 + 3 \omega^2 \tau^2 -2) + \omega \tau (\omega^2 \tau^2 + 5)}{(1+\omega^2\tau^2)^3
    }, \\
     \nonumber\sigma^{xyz}_\text{dc} &=(\sigma^{yxz}_\text{dc})^* \approx -i\frac{\eta e^3 \omega_B^4 \omega \tau^4}{24\pi^2\mu^2}\frac{\omega^2 \tau^2 + 5}{(1+\omega^2\tau^2)^3}, \\ \nonumber\sigma^{zzz}_\text{dc} & \approx \frac{3\eta e^3 \omega_B^6 \tau^2 \hbar^3}{160\pi^2\mu^5}\frac{1}{1+\omega^2\tau^2}.
\end{align}
When deriving this result, we also assumed that $\omega \mu^2 \tau^3 \gg \hbar^2$ for simplicity, while more general expression up to order $\mathcal{O}(\omega_B^6)$ is presented in Appendix~\ref{appendix-D}.  Alternatively, it can  be viewed as the expansion in powers of $1/\mu$ at $\mu \to \infty$.

Equation~\eqref{semi-mu-limit} is consistent with the result obtained in Ref.~\cite{BednikKozii2024}, where the Boltzmann equations~\eqref{Diff_f2} were solved perturbatively in $\omega_B$ to order $\mathcal{O}(\omega_B^2)$. The exact solution obtained here furthermore allows us to systematically generate higher-order terms in $\omega_B$, which provide the leading contributions to several additional conductivity components. The validity of the semiclassical approach for such higher-order contributions, however, is unclear. For example, the resulting expressions do not correctly reproduce the nonzero $\sigma^{xyz}$ component in the $B=0$ limit obtained from the Kubo formula in Appendix~\ref{AppB} and given by Eq.~\eqref{Eq:xyzB=0}. As we discuss below, similar discrepancies between the Boltzmann and Kubo-formula approaches arise in several other limits as well.

A related subtlety is that the straightforward solution of the Boltzmann equation reveals a $B$-independent contribution to the second-order photocurrent, which we omitted from Eq.~\eqref{semi-mu-limit}. It can be identified with the Berry-curvature-dipole contribution~\cite{SodemannFu2015}:
\be  \label{Eq:BCDcurrent}
\mathbf{j}_{\text{BCD}} = \frac{i \eta e^3 }{6\pi^2\hbar^2}\frac{\tau^2\omega}{1+\omega^2\tau^2}\mathbf{E}_\omega\times\mathbf{E}^*_\omega. 
\ee 
This term was obtained, for example, in Ref.~\cite{BednikKozii2024} and has recently been discussed in the context of the intraband CPGE~\cite{GolubIvchenko2026}. We emphasize, however, that this contribution is independent of the chemical potential and proportional to the topological charge of the Weyl node. Consequently, it vanishes upon summation over nodes of opposite chirality, at least within the simple model of identical Weyl nodes differing only in their position and the sign of their topological charge. Moreover, for a single Weyl node in the clean limit $\tau\to\infty$, neither the Fermi-golden-rule calculation~\cite{deJuan2017} nor the microscopic Kubo-formula approach~\cite{Avdoshkin2020} yields this contribution, as can be seen from Eq.~\eqref{Eq:xyzB=0} and Appendix~\ref{AppB}. This observation raises questions regarding its physical origin and relevance.

In the clean limit, $\tau\to\infty$, followed by the low-field limit $B\to0$ (equivalent to $\mu \to \infty$), we find, to leading order, the following contributions to each component:
\begin{align}  \label{semi-tauB-limit}
\nonumber\sigma^{xyz}_\text{dc} &=(\sigma^{yxz}_\text{dc})^* \approx -i\frac{\eta e^3 \omega_B^4}{24\pi^2\mu^2 \omega^3}, \\ 
   \sigma^{xxz}_\text{dc} &=\sigma^{yyz}_\text{dc} \approx -\frac{\eta e^3 \omega_B^6 \hbar}{24\pi^2\mu^3 \omega^4},
    \\
     \nonumber\sigma^{zzz}_\text{dc} & \approx \frac{3\eta e^3 \omega_B^6 \hbar^3}{160\pi^2\mu^5 \omega^2}.
\end{align}
Here we have also omitted terms analogous to Eq.~\eqref{Eq:BCDcurrent} and Eq.~\eqref{Eq:sigma-sigma}, which are independent of both $\mu$ and $\omega_B$. The remaining nonzero components arise only at higher order in either $1/\tau$ or $\omega_B^2$.

A comparison of Eqs.~\eqref{semi-mu-limit} and~\eqref{semi-tauB-limit} shows that the limits $B\to0$ and $\tau\to\infty$ commute for the components $\sigma^{xxz}_{\text{dc}}$, $\sigma^{xyz}_{\text{dc}}$, $\sigma^{zzz}_{\text{dc}}$, and their symmetry-related counterparts. In the case of $\sigma^{xxz}_{\text{dc}}$, this requires expanding Eq.~\eqref{semi-mu-limit} to order $\mathcal{O}(\omega_B^6)$ before taking the limit $\tau\to\infty$. For the remaining nonvanishing components, the two limits do not commute. This is consistent with the behavior observed in the Kubo-formula approach.

More importantly, neither expression reproduces the microscopic results obtained from the Kubo formula, Eqs.~\eqref{Eq:xxz_smallB} and~\eqref{Eq:xyz_smallB}, regardless of the order in which the limits are taken. Such inconsistencies between different theoretical approaches are a known issue in nonlinear transport. For example, Ref.~\cite{Guo2026} recently demonstrated that the ``intrinsic'' contribution to the second-order photocurrent in a static electric field, i.e., the part that is independent of the scattering rate, depends sensitively on the underlying dissipation mechanism (see also the references therein for a detailed discussion). This sensitivity persists for both the geometric contribution determined by the band-structure geometry and the kinetic contribution arising from the band dispersion. Similarly, the phenomenological treatment of scattering may underlie the discrepancies between the Kubo-formula and Boltzmann results obtained in the present work. A detailed investigation of this issue lies beyond the scope of the present paper and is left for future study.

In the clean low-frequency limit, $\tau \to \infty$ followed by $\omega \to 0$, we obtain 
\begin{align}  \label{semi-tauw-limit}
\nonumber\sigma^{xyz}_\text{dc} &=(\sigma^{yxz}_\text{dc})^* \approx -i\frac{2\eta e^3 \mu^2 \omega}{\pi^2\omega_B^4 \hbar^4}, \\ 
   \sigma^{xxz}_\text{dc} &=\sigma^{yyz}_\text{dc} \approx -\frac{2\eta e^3 \mu}{3\pi^2\omega_B^2 \hbar^3},
    \\
     \nonumber\sigma^{zzz}_\text{dc} & \approx \frac{3\eta e^3 \omega_B^6 \hbar^3}{160\pi^2\mu^5 \omega^2}.
\end{align}
The expressions for $\sigma^{xxz}_{\text{dc}}$ and $\sigma^{xyz}_{\text{dc}}$ reproduce parametrically the Kubo-formula results, Eqs.~\eqref{Eq:xxz_loww} and~\eqref{Eq:xyz_loww_lowB}, up to numerical prefactors, upon again neglecting terms of the form of Eq.~\eqref{Eq:BCDcurrent}. By contrast, the result for $\sigma^{zzz}_{\text{dc}}$ does not agree with the Kubo-formula calculation, despite appearing remarkably robust across the various limits considered above. This discrepancy may indicate a more general breakdown of the semiclassical Boltzmann approach at order $\mathcal{O}(1/\mu^5)$.

Finally, we reproduce the low-frequency divergences at $\omega \approx \hbar\omega_B^2/(2\mu)$ and demonstrate how finite single-particle scattering regularizes them, resulting in pronounced  peaks. To this end, we rewrite the current in terms of dimensionless variables
\begin{align}
    \lambda \equiv \hbar\omega_B/\mu, \hspace{0.3cm} \gamma \equiv \frac{\mu \omega}{\hbar\omega_B^2}, \hspace{0.3cm} \delta \equiv \omega\tau,
\end{align}
and expand it in the limit $\lambda\to0$ up to order $\mathcal{O}(1/\lambda^2)$, while keeping $\gamma$ and $\delta$ finite. Assuming further the clean limit $\delta\gg1$ and focusing on frequencies near the singularity, we obtain
\begin{align}
    \nonumber\sigma^{xxz}_\text{dc} & = \sigma^{yyz}_\text{dc} \approx -i\sigma^{xyz}_\text{dc} = i\sigma^{yxz}_\text{dc}
    \\ \nonumber
    \approx &
    - \frac{\eta e^3 \omega_B^2}{24 \pi^2 \hbar \mu} \frac1{\left(\omega - \frac{\hbar\omega_B^2}{2\mu}  \right)^2 + \frac1{\tau^2}},
    \\ \nonumber
    \sigma^{yzx}_\text{dc} & = -\sigma^{xzy}_\text{dc} = (\sigma^{zyx}_\text{dc})^* = -(\sigma^{zxy}_\text{dc})^* 
    \\ \nonumber
  \approx -i \sigma^{xzx}_\text{dc} & = -i\sigma^{yzy}_\text{dc} = -i (\sigma^{zxx}_\text{dc})^* = -i (\sigma^{zyy}_\text{dc})^* 
    \\ 
    \approx &  \frac{\eta e^3 \mu}{6\pi^2\hbar^3 \omega_B^2 \tau} \frac1{\omega - \frac{\hbar \omega_B^2}{2\mu} + \frac{i}{\tau}}.
    \label{semi-tau-limit2-v3}
\end{align}
The resulting expressions for $\sigma^{xxz}_{\text{dc}}=\sigma^{yyz}_{\text{dc}}$ and $\sigma^{xyz}_{\text{dc}}=-\sigma^{yxz}_{\text{dc}}$ are in good agreement with the microscopic Kubo-formula calculation up to numerical prefactors. The remaining components vanish in the clean limit, also in agreement with the Kubo-formula results. Additional details of the calculations presented in this section are provided in Appendix~\ref{appendix-D}.

\subsection{Second harmonic generation}
Now, using the exact solution for $f_1$ and $f_2$, we perform a similar analysis for the second harmonic generation. First, in the limit $B \to 0$ with the additional condition $\omega \mu^2 \tau^3 \gg \hbar^2$ (which is equivalent to $\mu \to \infty$), we find
\begin{align}     \label{semi2w-mu-limit3}
    \nonumber\sigma^{xzx}_{2\omega} &=\sigma^{yzy}_{2\omega}=\sigma^{zxx}_{2\omega}=\sigma^{zyy}_{2\omega} = -\frac{\sigma^{xxz}_{2\omega}}{2} = -\frac{\sigma^{yyz}_{2\omega}}{2}
    \\
    &\approx \frac{\eta e^3 \omega_B^2 \tau ^2}{48 \pi ^2 \hbar \mu  }\frac{ 2 -  3i\omega\tau  }{ (1-i  \omega \tau)^2 (1-2i \omega \tau)},\nonumber\\
      \sigma^{yzx}_{2\omega} &=\sigma^{zyx}_{2\omega}=-\sigma^{xzy}_{2\omega}=-\sigma^{zxy}_{2\omega}  
    \\
    &\approx i\frac{\eta e^3 \omega_B^4 \omega \tau^4}{96 \pi ^2  \mu^2  }\frac{ 2 -  3i\omega\tau  }{ (1-i  \omega \tau)^3 (1-2i \omega \tau)^2}, \nonumber \\ \sigma_{2\omega}^{zzz} &\approx \frac{3 \eta e^3 \omega_B^6 \tau^2 \hbar^3}{320\pi^2 \mu^5} \frac1{(1-i\omega \tau) (1-2i\omega \tau)}. \nonumber
\end{align}

In the clean low-field limit, $\tau \to \infty$ followed by $B \to 0$, we obtain 
\begin{align} \label{semi2w-tauB-limit}
    \sigma^{xzx}_{2\omega} &=\sigma^{yzy}_{2\omega}=\sigma^{zxx}_{2\omega}=\sigma^{zyy}_{2\omega} = -\frac{\sigma^{xxz}_{2\omega}}{2} = -\frac{\sigma^{yyz}_{2\omega}}{2} \nonumber
    \\ 
    &\approx -\frac{\eta e^3 \omega_B^2}{32 \pi ^2 \hbar \mu  \omega^2},\\
      \sigma^{yzx}_{2\omega} &=\sigma^{zyx}_{2\omega}=-\sigma^{xzy}_{2\omega}=-\sigma^{zxy}_{2\omega}  
    \approx i\frac{\eta e^3 \omega_B^4}{128 \pi ^2  \mu^2 \omega^3  }, \nonumber \\ \sigma_{2\omega}^{zzz} &\approx -\frac{3 \eta e^3 \omega_B^6 \hbar^3}{640\pi^2 \mu^5\omega^2} . \nonumber  
\end{align}
We see that Eq.~\eqref{semi2w-tauB-limit} can be obtained from Eq.~\eqref{semi2w-mu-limit3} if we additionally assume $\omega \tau \gg 1$. Consequently, we see that the two limits, $B\to 0$ and $\tau \to \infty$, commute for the SHG within the Boltzmann approach. Furthermore, comparing these results to the microscopic expressions obtained by the clean-limit Kubo formula  in Ref.~\cite{BednikKozii2024}, we find perfect agreement for all the components, except for $\sigma^{zzz}_{2\omega}$~\footnote{For components $\sigma^{yzx}_{2\omega}$ and other related by symmetry, we only consider the analytic part of the microscopic answer, which is proportional to $\omega_B^4$.}. This observation convinces us further in limited applicability of the Boltzmann approach, for this particular problem, at the order $\mathcal{O}(1/\mu^5)$.

In the clean low-frequency limit, $\tau \to \infty$ followed by $\omega \to 0$, we find
\begin{align}  \label{semi2w-tauomega-limit}
    \nonumber\sigma^{xzx}_{2\omega} &=\sigma^{yzy}_{2\omega}=\sigma^{zxx}_{2\omega}=\sigma^{zyy}_{2\omega} = -2\sigma^{xxz}_{2\omega} = -2\sigma^{yyz}_{2\omega}
    \\
    &\approx \frac{\eta e^3 \mu}{2 \pi ^2  \omega_B^2 \hbar^3},\\
      \sigma^{yzx}_{2\omega} &=\sigma^{zyx}_{2\omega}=-\sigma^{xzy}_{2\omega}=-\sigma^{zxy}_{2\omega}  
    \approx -i\frac{2\eta e^3 \mu^2 \omega}{\pi ^2  \hbar^4 \omega_B^4}, \nonumber \\ \sigma_{2\omega}^{zzz} &\approx -\frac{3 \eta e^3 \omega_B^6 \hbar^3}{640\pi^2 \mu^5\omega^2}. \nonumber
\end{align}
Here we again omitted contributions similar to Eq.~\eqref{Eq:BCDcurrent} that are independent of $\mu$, and assumed $\hbar \omega_B \ll \mu$. These results, except for $\sigma_{2\omega}^{zzz}$ component, are in perfect agreement with the Kubo-formula expressions from Ref.~\cite{BednikKozii2024}.

Finally, near the singularities $\omega \approx \hbar \omega_B^2/(2\mu)$ and $\omega \approx \hbar \omega_B^2/(4\mu)$, we find in the clean limit $\omega \tau \gg 1$: 
\begin{align}\label{semi2w-tau-limit2}
    &\sigma^{xxz}_{2\omega} =\sigma^{yyz}_{2\omega} \approx 
    \frac{\eta  e^3 }{16 \pi ^2 \hbar^2} \frac1{\omega-\frac{\hbar\omega_B^2}{2\mu} +\frac{i}{\tau}},
    \\ \nonumber 
   &\sigma^{zxx}_{2\omega}  = \sigma^{xzx}_{2\omega} = \sigma^{zyy}_{2\omega} = \sigma^{yzy}_{2\omega} 
    \\  \approx &
i\sigma^{zyx}_{2\omega}  = i\sigma^{yzx}_{2\omega} = -i\sigma^{zxy}_{2\omega} = -i\sigma^{xzy}_{2\omega} \nonumber 
    \\ 
     \approx 
   & -\frac{\eta  e^3 }{16 \pi ^2 \hbar^2}\left(\frac1{\omega-\frac{\hbar\omega_B^2}{4\mu} +\frac{i}{2\tau}} + \frac{i}{3\tau}\frac1{\left(\omega-\frac{\hbar\omega_B^2}{4\mu} +\frac{i}{2\tau}\right)^2} \right).\nonumber
\end{align}
These expressions reveal both first- and second-order poles at $\omega = \hbar \omega_B^2/4\mu$. By comparing them numerically with the Kubo-formula results of Ref.~\cite{BednikKozii2024}, we find good agreement provided that $1/\tau$ is identified with the phenomenologically introduced small imaginary part of the light frequency. The agreement becomes quantitatively more accurate, however, if we {\it neglect} the second-order pole, which is likely not captured by the clean-limit Kubo formula. While this discrepancy may be related to the additional factor of $1/\tau$ appearing in this term, which formally vanishes in the clean limit, we do not currently have a microscopic understanding of its origin. The details of the calculation are presented in Appendix~\ref{appendix-D}.

\section{Discussion and Conclusion}
\label{Conclusion}

In summary, we studied the shift current in the presence of a magnetic field for an ideal model of a clean, noninteracting Weyl node. Using the Kubo formula, we derived a general expression for the photoconductivity valid for arbitrary light frequency, magnetic-field strength, and Fermi energy. We identified several resonant frequencies at which the shift-current response diverges within the clean, noninteracting model. While these divergences are expected to be cut off by finite scattering in realistic systems, our results demonstrate that magnetic-field tuning can strongly enhance the photocurrent, thereby opening new opportunities for nonlinear magneto-optical phenomena. We further complemented the microscopic analysis with an exact solution of the Boltzmann equation, treating the magnetic field nonperturbatively. This approach enabled us to explore a broad range of semiclassical regimes, reproduce the low-frequency resonances obtained microscopically, and clarify the role of finite scattering in shaping the nonlinear response.

The main open question raised by our study is the inconsistency between the photocurrents obtained by different theoretical approaches in the clean limit across many regimes. As discussed above, this discrepancy may originate from the phenomenological treatment of scattering processes. Recent work has shown that the microscopic details of the dissipation mechanism can play a crucial role in determining the shift current, even when the resulting response is independent of the scattering rate and therefore appears entirely intrinsic~\cite{Guo2026}. While the primary focus of the present work was the effect of a magnetic field, a comprehensive microscopic theory of nonlinear photocurrents should account for relaxation processes on equal footing with the coherent dynamics. In particular, the shift current is known to receive important contributions associated with electron-hole recombination as well as intraband relaxation due to impurity and phonon scattering~\cite{Belinicheretal1982,ZhuAlexandradinata2024}. The Green's-function formalism employed in this work provides a natural framework for investigating the interplay between magnetic fields and such microscopic relaxation mechanisms.

Another open question concerns the fate and physical significance of the intraband contributions that are independent of the chemical potential, such as those given by Eqs.~\eqref{Eq:sigma-sigma} and~\eqref{Eq:BCDcurrent}. Within our simple model, these contributions vanish upon summation over Weyl nodes of opposite chirality. It is nevertheless natural to ask whether they can survive in more realistic systems. One possibility is to consider tilted Weyl nodes, which need not contribute equally to the photocurrent. Another route is provided by topological semimetals hosting nodes with different topological structures while maintaining an overall vanishing topological charge in the Brillouin zone. A notable example is CoSi, which in the absence of spin-orbit coupling hosts a threefold fermion and a fourfold (double-Weyl) fermion at different energies~\cite{RaoLi2019,NiWang2021}.

At the same time, the microscopic origin of these contributions remains unclear. In the Boltzmann approach, they appear already in the absence of a magnetic field and are associated with the Berry-curvature-dipole contribution. Within the Kubo formalism, however, a related term emerges only in the limit of a vanishing but finite magnetic field and is characterized by a different numerical prefactor. This discrepancy suggests that the physical origin of these contributions, and their relation to microscopic relaxation processes, deserves further investigation.

\section{Acknowledgments}
We thank Grigory Bednik and Maxim Dzero for collaboration on related projects. This work was supported by the Pittsburgh Quantum Institute Community Collaboration Award.

\appendix
\begin{widetext}
\section{Photocurrent at zero magnetic field \label{AppB}}
In this appendix, we derive an exact expression for the intrinsic contribution to the second-order conductivity tensor of an ideal clean Weyl node in the absence of a magnetic field, $B=0$. As shown in Refs.~\cite{Avdoshkin2020,BednikKozii2024} using the Kubo formula, the corresponding susceptibility tensor in Matsubara frequencies has the form
\begin{equation} \label{AppEq:B=0exact}
    \chi^{\alpha\beta\gamma}(i\omega_1,i\omega_2) = \frac{\eta e^3 \varepsilon^{\alpha\beta\gamma}}{48\pi^2 \hbar^2}\frac{\Omega^3(\omega_1-\omega_2)\ln{(4\mu^2+\hbar^2\Omega^2)}+\omega_2^3(\omega_1+\Omega)\ln{(4\mu^2+\hbar^2\omega_2^2)}-\omega_1^3(\omega_2+\Omega)\ln{(4\mu^2+\hbar^2\omega_1^2)}}{\omega_1\omega_2\Omega},
\end{equation}
where $\varepsilon^{\alpha\beta\gamma}$ is the fully antisymmetric Levi-Civita tensor and $\Omega = \omega_1 + \omega_2$. We see from this expression that $\chi^{\beta\alpha\gamma}(i\omega_2,i\omega_1) = \chi^{\alpha\beta\gamma}(i\omega_1,i\omega_2)$, so the conductivity tensor according to Eq.~\eqref{eq-sigma_xi} is given by $\sigma^{\alpha\beta\gamma}(i\omega_1,i\omega_2) = 2 \chi^{\alpha\beta\gamma}(i\omega_1,i\omega_2)/\omega_1 \omega_2$.


To obtain the physical photoconductivity at light frequency $\omega$, we analytically continue the Matsubara frequencies as
$i\omega_1 \to \omega + x\Omega + i0, \quad
i\omega_2 \to -\omega + (1-x)\Omega + i0,$
with $x\in[0,1]$, and subsequently expand the result for $\Omega\to0$ up to order $\mathcal{O}(\Omega^0)$. This yields
\be  
\sigma^{\alpha \beta \gamma}(\omega) = \left[  \frac{\eta e^3 \text{sign}(\omega) \Theta(\hbar^2 \omega^2 - 4\mu^2)}{12 \pi \hbar^2 \Omega} + \frac{i\eta e^3 \omega}{12 \pi^2 (\hbar^2 \omega^2 - 4\mu^2)}\right] \varepsilon^{\alpha \beta \gamma}.
\ee 
The first term in this expression reproduces the injection current quantization~\cite{deJuan2017,Avdoshkin2020}, while the second term describes the intrinsic shift current of a Weyl node. We note that this result does not depend on the specific choice of $x$.

Finally, we see from Eq.~\eqref{AppEq:B=0exact} that all the second harmonic generation components at zero field vanish after analytical continuation $i\omega_{1,2} \to \omega + i0$.


\section{Analytic continuation of the correlation function $\chi^{\alpha\beta\gamma}(i\omega_1,i\omega_2)$ \label{AppA}}
To extract physical shift current, we analytically continue Eq.~\eqref{xi2} as $i\omega_1\rightarrow \omega+\Omega/2+i0$, $i\omega_2\rightarrow-\omega+\Omega/2+i0$.
Expanding the result in power of $1/\Omega$ at  $\Omega\rightarrow 0$ and picking the part independent of $\Omega$, we find

\begin{align}  \label{shiftcurrent_correlation}
  &\chi^{\alpha\beta\gamma}\left(\omega + \frac{\Omega}2+i0,-\omega + \frac{\Omega}2+i0\right) \to    \chi^{\alpha\beta\gamma}(\omega)
    = (ev_f)^3\frac{\eta e B}{2\pi \hbar} \text{v.p.}\int\frac{dk_z}{2\pi} \sum_{n_1,n_2,n_3} Z^{\alpha\beta\gamma}_{n_1,n_2,n_3}\times  
    \\
    &\times\left\{ \left[\Theta(\varepsilon_{n_2})-\Theta(\varepsilon_{n_1})\right]\delta_{n_2,n_3}
    \frac{d}{d\xi}\left.\frac{1}{\hbar\omega+E_{n_1}-E_{n_2}+\xi}\right\vert_{\xi\rightarrow 0}  \right.
    \nonumber\\
    &-\left[\Theta(\varepsilon_{n_2})-\Theta(\varepsilon_{n_1})\right](1-\delta_{n_2,n_3})
    \left[
    \frac{1}{(\hbar\omega+E_{n_1}-E_{n_2})(\hbar\omega+E_{n_1}-E_{n_3})} 
    + \frac{i\pi\delta(\hbar\omega+E_{n_1}-E_{n_3})}{E_{n_2}-E_{n_3}}+ \right.
    \nonumber\\
    & \left.\left. + \frac{i\pi\delta(\hbar\omega+E_{n_1}-E_{n_2})}{E_{n_2}-E_{n_3}}
    \right] - 
    \left[\Theta(\varepsilon_{n_2})-\Theta(\varepsilon_{n_3})\right]
    \bigg[
    \frac{1}{(E_{n_2}-E_{n_3})(\hbar\omega+E_{n_1}-E_{n_3})}-\frac{i\pi\delta(\hbar\omega+E_{n_1}-E_{n_3})}{E_{n_2}-E_{n_3}}
    \bigg]
    \right\}, \nonumber
\end{align}
where $\text{v.p.}$ stands for the principal value integral. We use this expression as a starting point for calculating various components of the shift current. 


\section{Shift current \label{AppC}}
\subsection{$\sigma^{xxz}$ component}

To derive the conductivity components, we start with Eq.~\eqref{shiftcurrent_correlation}. The matrix element for $\sigma^{xxz}$ equals
\begin{align}
    Z^{xxz}_{n_1,n_2,n_3} =& 
    -\frac{E_0}{4E_{n_1}E_{n_2}^2}\delta_{n_2,n_3} \bigg[\delta_{|n_2|,|n_1|-1}(E_{n_2}+E_0)(E_{n_1}-E_0)+\delta_{|n_2|,|n_1|+1}(E_{n_2}-E_0)(E_{n_1}+E_0)\bigg] 
    \nonumber\\
    &-\frac{E_{n_2}^2-E_0^2}{4E_{n_1}E_{n_2}^2}\delta_{n_2,-n_3} \bigg[\delta_{|n_2|,|n_1|-1}(E_{n_1}-E_0)-\delta_{|n_2|,|n_1|+1}(E_{n_1}+E_0) \bigg] .
\end{align}

Plugging the above expression into Eq.~\eqref{shiftcurrent_correlation} and performing straightforward manipulations, we get
\begin{align}
     \chi^{xxz} &= (ev_f)^3\frac{\eta e B}{2\pi \hbar} \int\frac{dk_z}{2\pi} \sum_{n_1,n_2} \frac{\Theta(\varepsilon_{n_2})-\Theta(\varepsilon_{n_1})}{4 E_{n_1}E^2_{n_2}}\times
    \nonumber\\
    &\times\bigg\{ - E_0
    \left.\frac{d}{d\zeta}\frac{1}{E_{n_1}-E_{n_2}+\zeta}\right\vert_{\zeta\rightarrow\hbar\omega}
    \bigg[\delta_{|n_2|,|n_1|-1}(E_{n_2}+E_0)(E_{n_1}-E_0)+\delta_{|n_2|,|n_1|+1}(E_{n_2}-E_0)(E_{n_1}+E_0)\bigg]+
    \nonumber\\
    &+\frac{E_{n_2}^2-E_0^2}{E_{n_2}(\hbar\omega+E_{n_1}-E_{n_2})}
    \bigg[\delta_{|n_2|,|n_1|-1}(E_{n_1}-E_0)-\delta_{|n_2|,|n_1|+1}(E_{n_1}+E_0) \bigg]\bigg\}.
\end{align}

For convenience, we split this expression into the quantum contribution $\chi_q$ that includes the $0$-th Landau level and the nonquantum contribution $\chi_{nq}$ that includes terms with $n_{1,2}\neq 0$, such that $\chi^{xxz} = \chi_q^{xxz} + \chi_{nq}^{xxz}$. We write then
\begin{align}
    \chi^{xxz}_{nq} &= (ev_f)^3\frac{\eta e B}{2\pi \hbar} \text{v.p.}\int\frac{dk_z}{2\pi} \sum_{n_1,n_2\neq 0}[\Theta(\varepsilon_{n_2})-\Theta(\varepsilon_{n_1})](\delta_{|n_2|,|n_1|-1}-\delta_{|n_2|,|n_1|+1}\big)\times
    \nonumber\\&
    \times \bigg\{ 
    \frac{d}{d\zeta}\left.\frac{E_0^2(E_{n_2}-E_{n_1})}{4E_{n_1}E_{n_2}^2(E_{n_1}-E_{n_2}+\zeta)}\right\vert_{\zeta\rightarrow\hbar\omega}+
    \frac{E_{n_2}^2-E_0^2}{4E_{n_2}^3(\hbar\omega+E_{n_1}-E_{n_2})}
    \bigg\},
\end{align}
where v.p. refers to the Cauchy principal value integral. For the quantum contribution, we find 
\begin{align}
\chi_{q}^{xxz} &= (ev_f)^3 \frac{\eta e B}{2 \pi \hbar}\hspace{0.1cm}\text{v.p.}\int \frac{dk_z}{2\pi}\sum_{n} \delta_{|n|,1} [\Theta(\varepsilon_{n})-\Theta(\varepsilon_{0})]\frac{(E_n - E_0)}{2E_n} \times 
\nonumber\\&
\times\left[
    \frac{d}{d\zeta} \left.\left(\frac{1}{\zeta+E_{n}-E_{0}} -
    \frac{E_0}{E_n(\zeta+E_0-E_{n})} \right)\right\vert_{\zeta\rightarrow \hbar\omega}
    -\frac{(E_n+E_0)}{E_n^2(\hbar\omega+E_0-E_{n})} 
\right]. 
\end{align}

To simplify the expression for conductivity significantly, we add  the second term from Eq.~\eqref{eq-sigma_xi}, $\chi^{\beta \alpha \gamma}(i\omega_2,i\omega_1)$, and perform analytical continuation. As a result, we find that $\sigma^{xxz}(\omega)$ is an even function of frequency $\omega$ and odd function of chirality $\eta$ and chemical potential $\mu$. Focusing on the case $\mu > 0$ for definiteness, the resulting expression after some simplification takes the form 

\begin{align}
        \sigma^{xxz}(\omega) &= \frac{\eta e^4 v_f^2 B}{2\pi^2\hbar^3\omega^2}\text{v.p.}\int_0^\infty dk \sum_{n=0}^\infty [\Theta(\varepsilon_{n+1})-\Theta(\varepsilon_{n})]
        \left\{
        \frac{d}{d\zeta}\left.\frac{2 \zeta ^3 \left(\zeta ^4-2 \zeta ^2 \omega_B^2-4 \zeta ^2 n \omega_B^2+\omega_B^4\right)}{\left(\zeta ^4-\omega_B^4\right) \left(\zeta ^4-2(2n+1) \zeta ^2 \omega_B^2-4 \zeta ^2 k^2+\omega_B^4\right)}\right\vert_{\zeta\rightarrow\omega}+ \right.
        \nonumber\\&\left.
       +\frac{4 \omega ^2 \omega_B^2 \left[-(2 n+1) \omega ^6 +3 \omega ^4 \omega_B^2 -3 (2 n+1) \omega ^2 \omega_B^4+\omega_B^6\right]}{\left(\omega ^4-\omega_B^4\right)^2 \left(\omega ^4-4 k^2 \omega ^2-2 (2 n+1) \omega ^2 \omega_B^2+\omega_B^4\right)}
        \right\},
        \label{xxz-integral}
\end{align}
where we defined $k=k_z v_f$. A crucial subtlety here is that taking the derivative with respect to $\zeta$ before performing the integral leads to unphysical divergences. To regularize it, we first evaluate the integral over $k$ and only then take the derivative. This order of steps can be rigorously justified by keeping $\Omega$ small but finite and doing all the integrals before taking the limit $\Omega \to 0$ at the very end of the calculation. This procedure ensures a well-defined physically meaningful result given by Eq.~\eqref{xxz-sigma}, and the corresponding density plot is presented in Fig.~\ref{fig-xxz-2D-1}.

The resonant frequencies predicted by Eq.~\eqref{freq_w} are clearly seen in Fig.~\ref{fig-xxz-2D-1} as sharp red lines. The asymptotic expressions for $\sigma^{xxz}(\omega)$ near these resonances are given by (assuming that $\hbar \omega, \mu >0$)
\begin{align}
\sigma^{xxz}(\omega) &\approx \mp \frac{\eta e^3 \hbar \omega_B^2}{16 \pi^2  \mu \sqrt{\mu^2 + \hbar^2 \omega_B^2} \left( \sqrt{\mu^2 + \hbar^2 \omega_B^2} \pm \mu  \right)} \frac1{\omega - \omega_1^{\pm}}\sum_{n=0}^{\infty} k_n, \qquad &&\text{near} \quad \hbar\omega_1^{\pm} = \pm \mu + \sqrt{\mu^2+\hbar^2\omega_B^2}, \nonumber
\\
\sigma^{xxz}(\omega) &\approx \pm \frac{\eta e^3 \hbar \omega_B^2}{16 \pi^2  \mu \sqrt{\mu^2 - \hbar^2 \omega_B^2} \left(\mu  \pm \sqrt{\mu^2 - \hbar^2 \omega_B^2}   \right)} \frac1{\omega - \omega_2^{\pm}}\sum_{n=0}^{\infty} k_{n+1}, \qquad &&\text{near} \quad \hbar\omega_2^{\pm} = \mu \pm \sqrt{\mu^2-\hbar^2\omega_B^2}, 
\end{align}
where $k_n$ is given by Eq.~\eqref{definitions}.





\begin{figure}[h]
    \includegraphics[width=0.7\linewidth]{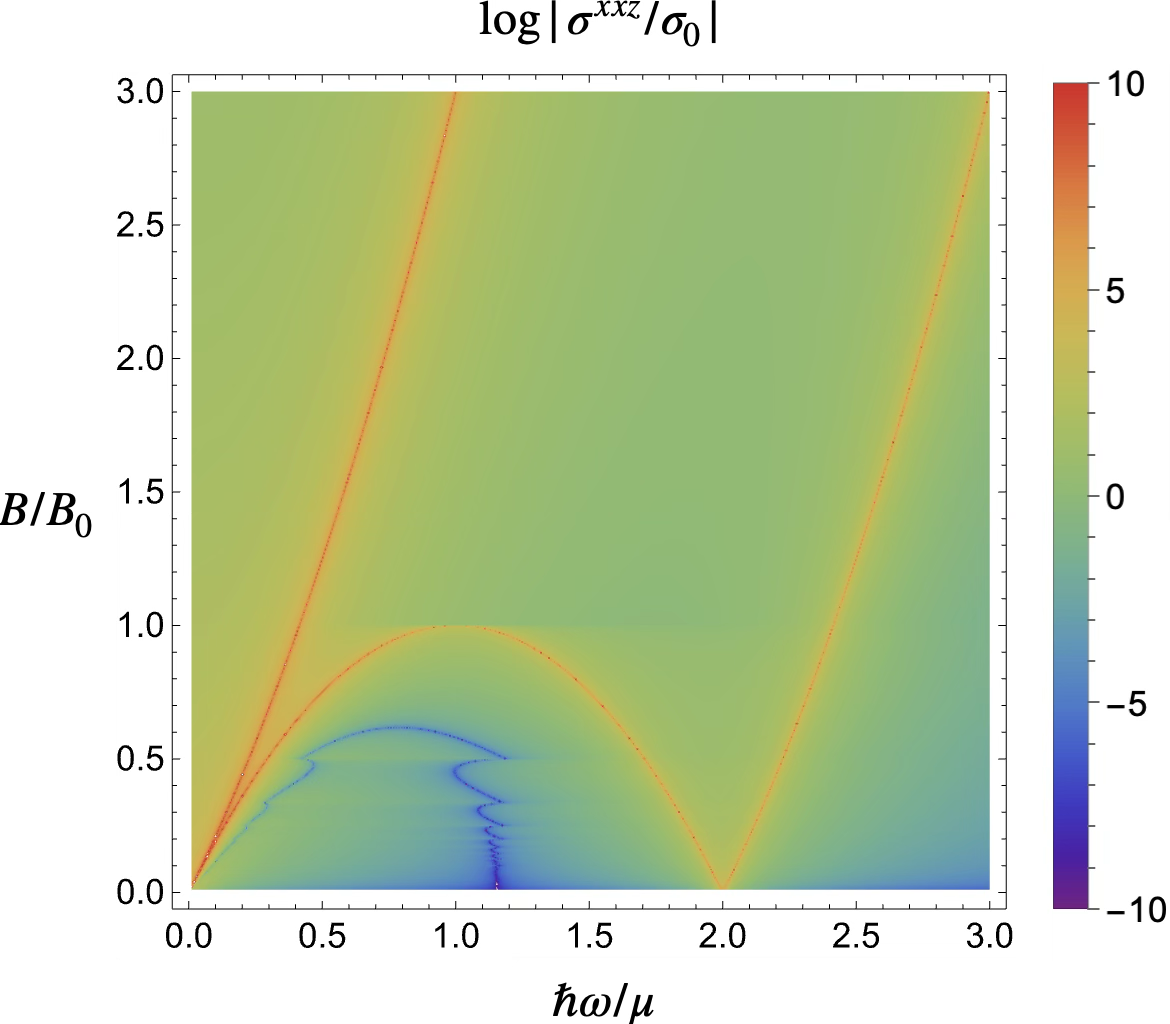}
    \caption{Density plot of $\log|\sigma^{xxz}/\sigma_0|$ as a function of frequency and magnetic field, Eq.~\eqref{xxz-sigma}. The units are $\sigma_0=\eta e^3/16\pi^2\hbar\mu$  and $B_0=\mu^2/2e\hbar v_f^2.$ We
    keep small but finite imaginary part of frequency to mimic the effect of a finite scattering rate. }
    \label{fig-xxz-2D-1}
\end{figure}


\begin{figure}[!h]
    \includegraphics[width=0.7\linewidth]{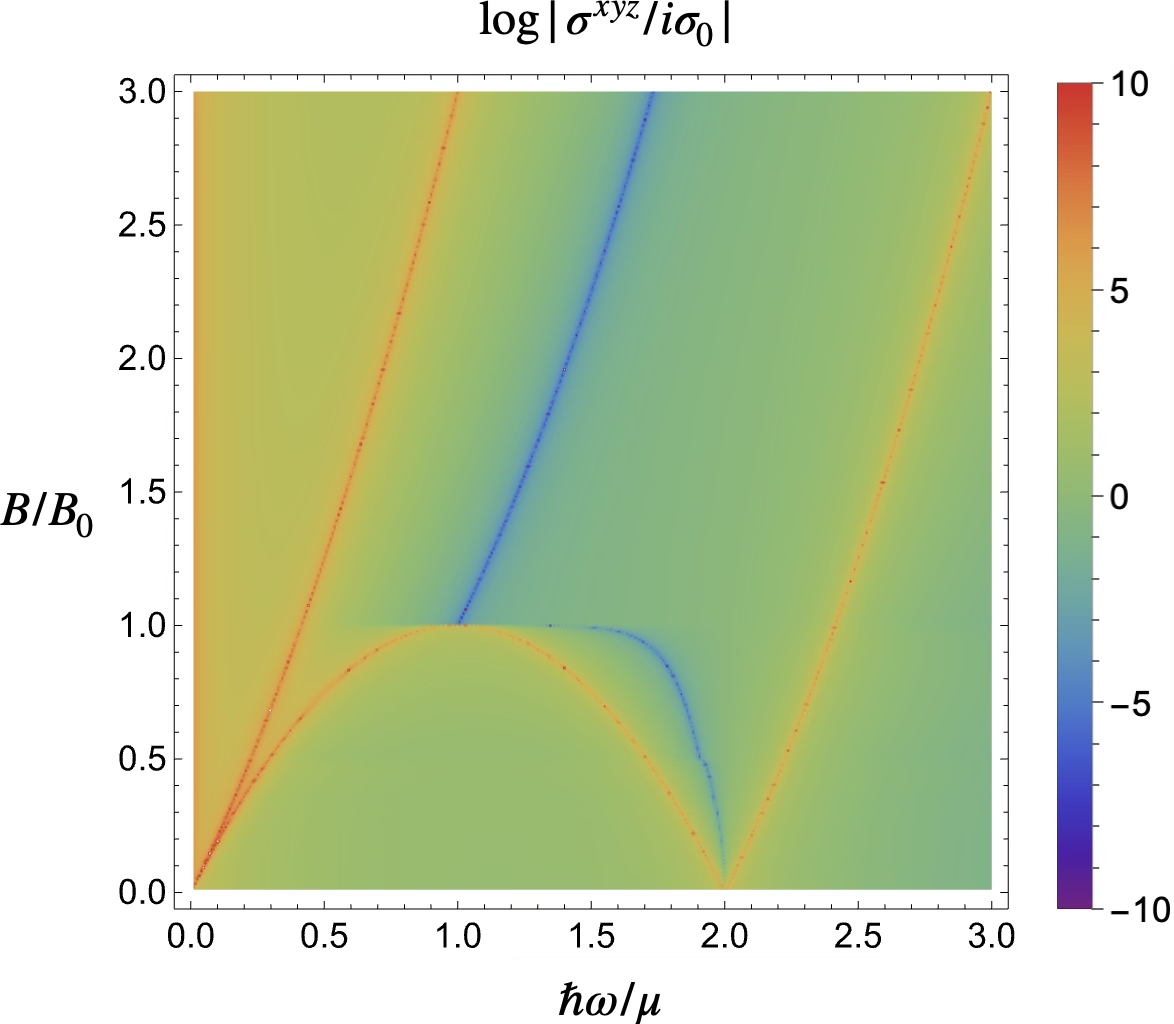}
    \caption{Density plot of $\log|\sigma^{xyz}/i\sigma_0|$ as a function of frequency and magnetic field, Eq.~\eqref{xyz-sigma}. The same units as in Fig.~\ref{fig-xxz-2D-1}.}
    \label{fig-xyz-2D-1}
\end{figure}

\subsection{$\sigma^{xyz}$ component}
The relevant matrix element for this component equals
\begin{align}
    Z^{xyz}_{n_1,n_2,n_3} &= 
    \frac{i E_0}{4E_{n_1}E_{n_2}^2}\delta_{n_2,n_3} \left[\delta_{|n_2|,|n_1|-1}(E_{n_2}+E_0)(E_{n_1}-E_0)-\delta_{|n_2|,|n_1|+1}(E_{n_2}-E_0)(E_{n_1}+E_0)\right] + 
    \nonumber\\&
    +\frac{i(E_{n_2}^2-E_0^2)}{4E_{n_1}E_{n_2}^2}\delta_{n_2,-n_3} \left[\delta_{|n_2|,|n_1|-1}(E_{n_1}-E_0)+\delta_{|n_2|,|n_1|+1}(E_{n_1}+E_0) \right].
\end{align}

Using this expression in Eq. \eqref{shiftcurrent_correlation}, we find
\begin{align}
        \chi^{xyz} &= (ev_f)^3\frac{\eta e B}{2\pi \hbar} \text{v.p.}\int\frac{dk_z}{2\pi} \sum_{n_1,n_2}  [\Theta(\varepsilon_{n_2})-\Theta(\varepsilon_{n_1})]\frac{i}{4E_{n_1}E_{n_2}^2}\times
        \nonumber\\&
        \times\bigg\{\frac{d}{d\zeta}\left.\frac{E_0}{E_{n_1}-E_{n_2}+\zeta}\right\vert_{\zeta\rightarrow\hbar\omega}\bigg[\delta_{|n_2|,|n_1|-1}(E_{n_2}+E_0)(E_{n_1}-E_0)-\delta_{|n_2|,|n_1|+1}(E_{n_2}-E_0)(E_{n_1}+E_0))\bigg]
        \nonumber\\&
        -\frac{(E_{n_2}^2-E_0^2)}{E_{n_2}(\hbar\omega+E_{n_1}-E_{n_2})}
        \bigg[\delta_{|n_2|,|n_1|-1}(E_{n_1}-E_0)+\delta_{|n_2|,|n_1|+1}(E_{n_1}+E_0)) \bigg]
        \bigg\}.
\end{align}

Upon adding the second contribution from Eq.~\eqref{eq-sigma_xi}, $\chi^{\beta \alpha \gamma}(i\omega_2,i\omega_1)$, and noting that $Z^{xyz}_{n_1,n_2,n_3}=-Z^{yxz}_{n_1,n_2,n_3}$, we perform analytical continuation and find that $\sigma^{xyz}(\omega)$ is an odd function of frequency $\omega$ and chirality $\eta$ and an even function of the chemical potential $\mu$. Focusing again on the case $\mu > 0$ for definiteness, we find after some simplification 
\begin{align}
        \sigma^{xyz}(\omega) &= i\frac{(ev_f)^3}{\omega^2}\frac{\eta e B}{\pi \hbar} \text{v.p.}\int_0^\infty\frac{dk_z}{2\pi} \sum_{n=0}^\infty
        \left\{\Theta(\varepsilon_{n})\left(
        \frac{E_0^2}{E_{n+1}^2E_{n}}\frac{d}{d\zeta}\left.\frac{\zeta^2(E_{n}^2+3E_{n+1}^2)-(E_n^2-E_{n+1}^2)^2}{\zeta^4-2\zeta^2(E_n^2+E_{n+1}^2)+(E_n^2-E_{n+1}^2)^2}\right\vert_{\zeta\rightarrow\hbar\omega} \right.\right.
        \nonumber\\&\left.
        -\frac{\hbar\omega}{E_n^3E_{n+1}^2}\frac{(\hbar\omega)^2E_{n+1}^2(E_n^2-E_0^2)+(E_{n+1}^2-E_n^2)(-E_{n+1}^2E_n^2+E_0^2E_{n+1}^2+2E_0^2E_n^2)}{(\hbar\omega)^4-2(\hbar\omega)^2(E_n^2+E_{n+1}^2)+(E_n^2-E_{n+1}^2)^2}\right)
        \nonumber\\&
        +\Theta(\varepsilon_{n+1})\left(
        \frac{E_0^2}{E_{n+1}E_{n}^2}\frac{d}{d\zeta}\left.\frac{\zeta^2(E_{n+1}^2+3E_{n}^2)-(E_n^2-E_{n+1}^2)^2}{\zeta^4-2\zeta^2(E_n^2+E_{n+1}^2)+(E_n^2-E_{n+1}^2)^2}\right\vert_{\zeta\rightarrow\hbar\omega}
        \right.\nonumber\\& \left.\left.
        -\frac{\hbar\omega}{E_n^2E_{n+1}^3}\frac{(\hbar\omega)^2E_{n}^2(E_{n+1}^2-E_0^2)+(E_{n}^2-E_{n+1}^2)(-E_{n+1}^2E_n^2+E_0^2E_{n}^2+2E_0^2E_{n+1}^2)}{(\hbar\omega)^4-2(\hbar\omega)^2(E_n^2+E_{n+1}^2)+(E_n^2-E_{n+1}^2)^2}\right)
        \right\}.
        \label{xyz-integral}
 \end{align}
 Performing integration over $k_z$ and simplifying the result, we end up with Eq.~\eqref{xyz-sigma}, which is also visualized in Fig.~\ref{fig-xyz-2D-1}.

The resonant frequencies of $\sigma^{xyz}(\omega)$ are given by the same Eq.~\eqref{freq_w} and can be clearly seen in Fig.~\ref{fig-xyz-2D-1} as sharp red lines. In their vicinity, we find

\begin{align}
\sigma^{xyz}(\omega) \approx &i \frac{\eta e^3 \hbar \omega_B^2}{16 \pi^2  \mu \sqrt{\mu^2 + \hbar^2 \omega_B^2} \left( \sqrt{\mu^2 + \hbar^2 \omega_B^2} \pm \mu  \right)} \frac1{\omega - \omega_1^{\pm}}\sum_{n=0}^{\infty} k_n, \qquad &&\text{near} \quad \hbar\omega_1^{\pm} = \pm \mu + \sqrt{\mu^2+\hbar^2\omega_B^2}, \nonumber
\\
\sigma^{xyz}(\omega) \approx \pm &i\frac{\eta e^3 \hbar \omega_B^2}{16 \pi^2  \mu \sqrt{\mu^2 - \hbar^2 \omega_B^2} \left(\mu  \pm \sqrt{\mu^2 - \hbar^2 \omega_B^2}   \right)} \frac1{\omega - \omega_2^{\pm}}\sum_{n=0}^{\infty} k_{n+1}, \qquad &&\text{near} \quad \hbar\omega_2^{\pm} = \mu \pm \sqrt{\mu^2-\hbar^2\omega_B^2}. 
\end{align}


\section{Semiclassical Description \label{appendix-D}}
We start our semiclassical Boltzmann analysis with the equations of motion for electrons in a periodic lattice potential:
\begin{align}
    \hbar\dot{\mathbf{r}} &= \mathbf{\nabla_k}\varepsilon_\mathbf{k} - \hbar\dot{\mathbf{k}}\times\boldsymbol{\Omega}_\mathbf{k},
    \nonumber\\
    \hbar\dot{\mathbf{k}} &= -e\mathbf{E} - e \dot{\mathbf{r}}\times\mathbf{B},
    \label{SM:motion_1}
\end{align}
where $\mathbf{\Omega_k}$ is the Berry curvature and $\varepsilon_\mathbf{k}$ is the modified energy in the presence of magnetic field given by
\begin{align}
    \mathbf{\Omega_k} &= i\bra{\nabla_\mathbf{k} u_\mathbf{k}}\times \ket{\nabla_\mathbf{k} u_\mathbf{k}},
    \nonumber\\ \varepsilon_\mathbf{k} &= \varepsilon^0_\mathbf{k} - \mathbf{m_k}\cdot\mathbf{B}.
\end{align}
Here, $\varepsilon_\mathbf{k}^0$ is the bare band energy at zero field, $H_\mathbf{k}|u_\mathbf{k}\rangle=\varepsilon_\mathbf{k}^0|u_\mathbf{k}\rangle$, and $\mathbf{m}_\mathbf{k}$ is the orbital magnetic moment of the electron wavepacket:
\begin{align}
    \mathbf{m_k} = -i\frac{e}{2\hbar}\bra{\nabla_\mathbf{k}u_\mathbf{k}}\times (H_\mathbf{k}-\varepsilon_\mathbf{k}^0)\ket{\nabla_\mathbf{k}u_\mathbf{k}}.
\end{align}
Equations~\eqref{SM:motion_1} can be solved to get
\begin{align}
    \dot{\mathbf{r}} &= \frac{1}{\hbar D_{\mathbf{k}}} \left\{ \nabla_{\mathbf{k}} \varepsilon_{\mathbf{k}} + e \mathbf{E} \times \bm{\Omega}_{\mathbf{k}} + \frac{e}{\hbar} \mathbf{B} (\bm{\Omega}_{\mathbf{k}} \cdot \nabla_{\mathbf{k}} \varepsilon_{\mathbf{k}}) \right\}, 
    \nonumber\\
    \dot{\mathbf{k}} &= \frac{1}{\hbar D_{\mathbf{k}}} \left\{ -e \mathbf{E} - \frac{e}{\hbar} \nabla_{\mathbf{k}} \varepsilon_{\mathbf{k}} \times \mathbf{B} - \frac{e^2}{\hbar} (\mathbf{E} \cdot \mathbf{B}) \bm{\Omega}_{\mathbf{k}} \right\},
\end{align}
where we introduced the phase space volume correction factor $D_\mathbf{k}=1+(e/\hbar)(\mathbf{B}\cdot\mathbf{\Omega_k})$~\cite{XiaoDOS2005}. 

The electric current equals 
\begin{align}
    \mathbf{j} = -e\sum_\zeta\int \frac{d^3k}{(2\pi)^3}D_\mathbf{k}\dot{\mathbf{r}}f = -\frac{e}{\hbar} \sum_\zeta\int \frac{d^3k}{(2\pi)^3}\bigg[ \nabla_\mathbf{k}\varepsilon_\mathbf{k} + e\mathbf{E}\times\mathbf{\Omega}_\mathbf{k}+\frac{e}{\hbar}\mathbf{B}(\nabla_\mathbf{k}\varepsilon_\mathbf{k}\cdot\mathbf{\Omega_k})\bigg]f, 
    \label{current}
\end{align}
where $\sum_\zeta$ stands for the summation over different bands (we suppress index $\zeta$ inside the integral) and $f$ is the distribution function that satisfies the kinetic Boltzmann equation. Assuming uniform system and the relaxation time approximation, it takes form
\begin{align}
    \frac{\partial f}{\partial t}+\dot{\mathbf{k}}\cdot\nabla_\mathbf{k}f=-\frac{f-f_0}{\tau}.
    \label{kinetic}
\end{align}
The zero-temperature equilibrium Fermi-Dirac distribution is given by $f_0=\Theta(\mu-\varepsilon_\mathbf{k})$, and $\tau$ is the relaxation time. Assuming a monochromatic electric field,
\begin{align}
    \mathbf{E}(t)=\mathbf{E}_\omega e^{-i\omega t}+\mathbf{E}_\omega^* e^{i\omega t},
\end{align}
with $\mathbf{E}_\omega^* = \mathbf{E}_{-\omega}$, we can expand the steady-state distribution function as 
\begin{align}
    f(t) = f_0 +\delta f_0 + (f_1 e^{-i\omega t}+f_1^*e^{i\omega t})+(f_2 e^{-2i\omega t}+f_2^*e^{2i\omega t})+\ldots,
\end{align}
where $f_1\propto E=|\mathbf{E}_\omega|$ and $\delta f_0,f_2 \propto E^2$. Matching various harmonics in Eq.~\eqref{kinetic}, we arrive at 
\begin{align} \label{diffeqs}
    D_{\mathbf{k}} \left( i\omega - \frac{1}{\tau} \right) f_1 &= -\frac{e}{\hbar} \left[ \mathbf{E}_\omega + \frac{e}{\hbar} (\mathbf{E}_\omega \cdot \mathbf{B}) \boldsymbol{\Omega}_\mathbf{k} \right] \cdot \nabla_{\mathbf{k}} f_0 
    - \frac{e}{\hbar^2} \nabla_{\mathbf{k}} f_1 \cdot \nabla_{\mathbf{k}} \varepsilon_{\mathbf{k}} \times \mathbf{B}, \nonumber
    \\
    D_{\mathbf{k}}\left(2i\omega - \frac{1}{\tau}\right)f_2 &= -\frac{e}{\hbar} \left[\mathbf{E}_\omega + \frac{e}{\hbar}(\mathbf{E}_\omega \cdot \mathbf{B})\mathbf{\Omega_k} \right] \cdot \nabla_{\mathbf{k}} f_1
    - \frac{e}{\hbar^2} \nabla_{\mathbf{k}} f_2 \cdot \nabla_{\mathbf{k}} \varepsilon_{\mathbf{k}} \times \mathbf{B},
    \\
    - D_{\mathbf{k}} \frac{1}{\tau} \delta f_0 &= -\frac{e}{\hbar} \left[ \mathbf{E}_\omega^* + \frac{e}{\hbar} (\mathbf{E}_\omega^* \cdot \mathbf{B}) \boldsymbol{\Omega}_\mathbf{k} \right] \cdot \nabla_{\mathbf{k}} f_1 
    - \frac{e}{2\hbar^2} \nabla_{\mathbf{k}} \delta f_0 \cdot \nabla_{\mathbf{k}} \varepsilon_{\mathbf{k}} \times \mathbf{B} + \text{c.c.} \nonumber
\end{align}

The corresponding current components extracted from Eq.~\eqref{current} can then be written as
\begin{align}\label{jdc}
    \mathbf{j}_1^{\omega}&=-\frac{e}{\hbar} \sum_\zeta\int \frac{d^3k}{(2\pi)^3}\bigg\{ \left[
    \nabla_\mathbf{k}\varepsilon_\mathbf{k} + \frac{e}{\hbar}(\nabla_\mathbf{k}\varepsilon_\mathbf{k}\cdot\mathbf{\Omega_k})\mathbf{B}\right]f_1+
    e \mathbf{E}_\omega\times\mathbf{\Omega_k}f_0
    \bigg\},\nonumber
    \\ 
    \mathbf{j}_2^{2\omega} &= -\frac{e}{\hbar} \sum_\zeta \int \frac{d^3k}{(2\pi)^3}\bigg\{ \left[\nabla_{\mathbf{k}}\varepsilon_{\mathbf{k}}+\frac{e}{\hbar}(\nabla_{\mathbf{k}}\varepsilon_{\mathbf{k}}\cdot \mathbf{\Omega_k})\mathbf{B}\right]f_2 +e \mathbf{E}_{\omega}\times\mathbf{\Omega_k}f_1\bigg\},
    \\
     \mathbf{j}_2^{\text{dc}} &= -\frac{e}{\hbar} \sum_\zeta \int \frac{d^3k}{(2\pi)^3}\bigg\{ \left[\nabla_{\mathbf{k}}\varepsilon_{\mathbf{k}}+\frac{e}{\hbar}(\nabla_{\mathbf{k}}\varepsilon_{\mathbf{k}}\cdot \mathbf{\Omega_k})\mathbf{B}\right]\delta f_0 +e \mathbf{E}_{\omega}\times\mathbf{\Omega_k}f_1^*+e \mathbf{E}_{\omega}^*\times\mathbf{\Omega_k}f_1
    \bigg\}.\nonumber
\end{align}
The component $\mathbf{j}_1^{\omega} \propto E$ describes the ordinary linear response, while $\mathbf{j}_2^{\text{dc}}, \mathbf{j}_2^{2\omega} \propto E^2$ correspond to the photocurrent and the second harmonic generation, respectively.

While Eqs.~\eqref{diffeqs} are generically complicated partial differential equations which can only be solved numerically or perturbatively in $\mathbf{B}$~\cite{BednikKozii2024,Morimoto2016}, they admit {\it exact} analytical solution for an ideal Weyl node considered in this work. In this case, the band dispersion, the Berry curvature, and the orbital magnetic moment for a single node are given by
\begin{align}
    \varepsilon_\mathbf{k}^0=\zeta v_f \hbar k, \hspace{0.5cm} \mathbf{\Omega_k}=-\frac{\zeta\eta}{2k^3}\mathbf{k}, \hspace{0.5cm} \mathbf{m_k}=-\eta\frac{ev_f}{2k^2}\mathbf{k},
\end{align}
where $\eta=\pm 1$ is the chirality of the node, $\zeta=+1/-1$ refers to the conduction/valence band, and $v_f$ is the Fermi velocity in the vicinity of the node. Assuming $\mathbf{E}_\omega = (E_x, E_y, E_z)$ and $\mathbf{B}=(0, 0, B)$, the last term in Eqs.~\eqref{diffeqs}, which makes the problem complicated, simplifies considerably in spherical coordinates due to the high symmetry of the Weyl node:
\begin{align}
    \nabla_{\mathbf{k}} f \cdot \left(\nabla_{\mathbf{k}} \varepsilon_{\mathbf{k}} \times \mathbf{B}\right)
    =
    -\frac{B}{k}\,\zeta \hbar v_f\left(1-\eta\zeta\frac{B e}{\hbar k^2}\cos\theta\right)\partial_\phi f .
\end{align}
As it only involves the derivative with respect to the azimuthal angle $\phi$, the equations reduce to ordinary differential equations that treat $k$ and the polar angle $\theta$ as parameters and can be readily resolved. Requiring the solution to be periodic in $\phi$, we find for the first-order correction: 
\be  \label{SMEq:f1}
f_1 = \boldsymbol{\alpha} \cdot \mathbf{S} f'_0(\ve_\bk), \qquad \boldsymbol{\alpha} = (\alpha_1,\alpha_2,\alpha_3), \qquad \mathbf{S} = (\cos\phi,\sin\phi,1),
\ee 
where we introduced
\begin{align}  &\alpha_1 = \frac{\beta z_\omega E_x-E_y}{B(1+\beta^2 z_\omega^2)}\hbar k \sin\theta, \hspace{0.5cm}
    \alpha_2 = \frac{E_x+\beta z_\omega E_y}{B(1+\beta^2 z^2_\omega)}\hbar k \sin\theta, \hspace{0.5cm}
    \alpha_3 = \frac{ev_f\tau\zeta\left(1-\frac{\zeta\eta\omega_B^2}{2v_f^2 k^2}\cos\theta+\frac{\omega_B^4}{16v_f^4k^4}\right)}{z_\omega\left(1-\frac{\zeta\eta\omega_B^2}{4v_f^2 k^2}\cos\theta \right)}E_z\cos\theta, \nonumber \\ 
&\beta = \frac{2k v_f \zeta \left(1-\frac{\zeta\eta\omega_B^2 k_z}{4v_f^2k^3}\right)}{\omega_B^2 \tau\left(1-\frac{\zeta\eta\omega_B^2 k_z}{2v_f^2k^3}\right)}, \qquad z_\omega = 1 - i \omega \tau, \qquad \omega_B^2 = 2 e B v_f^2/\hbar, \qquad f_0' = \partial_\ve f_0.
\end{align}
In the Cartesian coordinates, it takes form 
\be  
f_1 = \left(\frac{\beta z_\omega E_x-E_y}{B(1+\beta^2 z_\omega^2)}\hbar k_x + \frac{E_x+\beta z_\omega E_y}{B(1+\beta^2 z_\omega^2)}\hbar k_y + \frac{ev_f\tau\zeta \left(1-\frac{\zeta\eta\omega_B^2 k_z}{2v_f^2 k^3}+\frac{\omega_B^4}{16v_f^4 k^4}\right)}{z_\omega k \left(1-\frac{\zeta\eta\omega_B^2 k_z}{4v_f^2 k^3}\right)}E_z k_z \right) f_0'(\ve_\bk).
\ee

The second-order corrections $f_2$ and $\delta f_0$ can be found analogously. After lengthy but straightforward calculation, we find: 
\begin{align}
    f_2 &= \frac{\hbar k f''_0(\ve_\bk)}{B} \boldsymbol{\alpha}\cdot\left[E_x\sin\theta\mathbf{I_1}(\beta z_{2\omega})+E_y\sin\theta\mathbf{I_2}(\beta z_{2\omega})+E_z\cos\theta\mathbf{I_3}(\beta z_{2\omega})\frac{1-\frac{\zeta\eta\omega_B^2}{2v_f^2 k^2}\cos\theta+\frac{\omega_B^4}{16v_f^4k^4}}{1-\frac{\zeta\eta\omega_B^2}{2v_f^2 k^2}\cos\theta} \right]+
    \nonumber\\
    &+ \frac{\zeta kf'_0(\ve_\bk)}{B v_f \left(1-\frac{\zeta\eta\omega_B^2}{2v_f^2 k^2}\cos\theta\right)}\left[
    E_x\left(
    \sin\theta\partial_k\boldsymbol{\alpha}\cdot\mathbf{I_1}(\beta z_{2\omega})+\frac{\cos\theta}{k}\partial_\theta\boldsymbol{\alpha}\cdot\mathbf{I_1}(\beta z_{2\omega})-\frac{1}{k\sin\theta}\boldsymbol{\alpha}\cdot\mathbf{I_5}(\beta z_{2\omega})
    \right)+ \right. 
    \nonumber\\
    &+E_y\left(
    \sin\theta\partial_k\boldsymbol{\alpha}\cdot\mathbf{I_2}(\beta z_{2\omega})+\frac{\cos\theta}{k}\partial_\theta\boldsymbol{\alpha}\cdot\mathbf{I_2}(\beta z_{2\omega})+\frac{1}{k\sin\theta}\boldsymbol{\alpha}\cdot\mathbf{I_4}(\beta z_{2\omega})
    \right)+
    \nonumber\\
    &\left.+E_z\left(
    \left(\cos\theta-\frac{\zeta\eta\omega_B^2}{4v_f^2 k^2}\right)\partial_k\boldsymbol{\alpha}\cdot\mathbf{I_3}(\beta z_{2\omega})-\frac{\sin\theta}{k}\partial_\theta\boldsymbol{\alpha}\cdot\mathbf{I_3}(\beta z_{2\omega})
    \right)
    \right],
\label{f2}
\end{align}
\begin{align}
    \delta f_0 &= \frac{\hbar k f''_0(\ve_\bk)}{B} \boldsymbol{\alpha}\cdot\left[E^*_x\sin\theta\mathbf{I_1}(\beta)+E^*_y\sin\theta\mathbf{I_2}(\beta)+E^*_z\cos\theta\mathbf{I_3}(\beta)\frac{1-\frac{\zeta\eta\omega_B^2}{2v_f^2 k^2}\cos\theta+\frac{\omega_B^4}{16v_f^4k^4}}{1-\frac{\zeta\eta\omega_B^2}{2v_f^2 k^2}\cos\theta} \right]+
    \nonumber\\
    &+ \frac{\zeta kf'_0(\ve_\bk)}{B v_f \left(1-\frac{\zeta\eta\omega_B^2}{2v_f^2 k^2}\cos\theta\right)}\left[
    E^*_x\left(
    \sin\theta\partial_k\boldsymbol{\alpha}\cdot\mathbf{I_1}(\beta)+\frac{\cos\theta}{k}\partial_\theta\boldsymbol{\alpha}\cdot\mathbf{I_1}(\beta)-\frac{1}{k\sin\theta}\boldsymbol{\alpha}\cdot\mathbf{I_5}(\beta)
    \right)+ \right. 
    \nonumber\\
    &+E^*_y\left(
    \sin\theta\partial_k\boldsymbol{\alpha}\cdot\mathbf{I_2}(\beta)+\frac{\cos\theta}{k}\partial_\theta\boldsymbol{\alpha}\cdot\mathbf{I_2}(\beta)+\frac{1}{k\sin\theta}\boldsymbol{\alpha}\cdot\mathbf{I_4}(\beta)
    \right)+
    \nonumber\\
    &\left.+E^*_z\left(
    \left(\cos\theta-\frac{\zeta\eta\omega_B^2}{4v_f^2 k^2}\right)\partial_k\boldsymbol{\alpha}\cdot\mathbf{I_3}(\beta)-\frac{\sin\theta}{k}\partial_\theta\boldsymbol{\alpha}\cdot\mathbf{I_3}(\beta)
    \right)
    \right] + \text{c.c.},
\label{deltaf0}
\end{align}
where $f_0'' = \partial^2_{\ve} f_0$, $z_{2\omega} = 1 - 2i\omega \tau$, and we defined
\begin{align}
    \mathbf{I_1}(x) &= \left(\frac{x^2 \cos (2 \phi )+x^2+2 x \sin (2 \phi )+4}{2 x^3+8 x},\frac{x \sin (2 \phi )-2 \cos (2 \phi )}{2 \left(x^2+4\right)},\frac{x \cos (\phi )+\sin (\phi )}{x^2+1}\right),
    \nonumber\\
    \mathbf{I_2}(x) &= \left(\frac{x \sin (2 \phi )-2 \cos (2 \phi )}{2 \left(x^2+4\right)},\frac{x^2 (-\cos (2 \phi ))+x^2-2 x \sin (2 \phi )+4}{2 x^3+8 x},\frac{x \sin (\phi )-\cos (\phi )}{x^2+1}\right),
    \nonumber\\
    \mathbf{I_3}(x) &= \left(\frac{x \cos (\phi )+\sin (\phi )}{x^2+1},\frac{x \sin (\phi )-\cos (\phi )}{x^2+1},\frac{1}{x}\right),
    \nonumber\\
    \mathbf{I_4}(x) &= \left(\frac{2 \cos (2 \phi )-x \sin (2 \phi )}{2 x^2+8},\frac{x^2 \cos (2 \phi )+x^2+2 x \sin (2 \phi )+4}{2 x^3+8 x},0\right),
    \nonumber\\
    \mathbf{I_5}(x) &= \left(-\frac{x^2 (-\cos (2 \phi ))+x^2-2 x \sin (2 \phi )+4}{2 x^3+8 x},\frac{x \sin (2 \phi )-2 \cos (2 \phi )}{2 \left(x^2+4\right)},0\right).
\end{align}
While the expressions for $f_2$ and $\delta f_0$ admit rewriting in Cartesian coordinates, we do not find them especially useful. In the limit of small magnetic fields, the solution~\eqref{SMEq:f1}-\eqref{deltaf0} reproduces the expressions from Ref.~\cite{BednikKozii2024} obtained by Taylor-expanding Eqs.~\eqref{diffeqs} up to linear order in $B$ and then solving them iteratively. 

Our exact solution~\eqref{SMEq:f1}-\eqref{deltaf0} can now be plugged into Eq.~\eqref{jdc} to calculate the components of nonlinear conductivity in various limits.

\subsection{Low-field limit $\omega_B\rightarrow 0$}
First, we calculate the components of the second-order current in the limit of the vanishing magnetic field $\omega_B \to 0$. Expanding the exact solution in Eq.~\eqref{jdc} in powers of $\omega_B^2$, we find for the photocurrent $\mathbf{j}_2^\text{dc}$:
\begin{align}
    \mathbf{j}_2^\text{dc} &\approx \frac{i \eta e^3 \omega \tau^2}{6\pi^2\hbar^2(1+\omega^2 \tau^2)}\mathbf{E}_\omega\times\mathbf{E}^*_\omega 
    \nonumber\\
    &+ \frac{\eta e^3\omega_B^2 \tau^2}{24\pi^2\hbar\mu(1+\omega^2\tau^2)^2}
    \left[
    \left(
        \begin{array}{c}
        2(E_x E_z^* + E_x^* E_z) \\
        2(E_y E_z^* + E_y^* E_z) \\
        -4(|E_x|^2 + |E_y|^2)
        \end{array}
    \right)+i\omega\tau(1-\omega^2\tau^2)
    \left(
        \begin{array}{c}
        E_x E_z^* - E_x^* E_z \\
        E_y E_z^* - E_y^* E_z \\
        0
        \end{array}
    \right)
    \right] \nonumber \\ &+ \frac{\eta e^3\omega_B^4 \omega  \tau^4}{48\pi^2\mu^2(1+\omega^2\tau^2)^3}
    \left[
    (5 + \omega^2 \tau^2)\left(
        \begin{array}{c}
        \omega \tau(E_y E_z^* + E_y^* E_z) \\
        -\omega \tau(E_x E_z^* + E_x^* E_z) \\
        -2i(E_x E_y^*-E_y E_x^*)
        \end{array}
    \right)+i(\omega^4 \tau^4 + 3 \omega^2 \tau^2 - 2)
    \left(
        \begin{array}{c}
        E_y E_z^* - E_y^* E_z \\
        E_z E_x^* - E_z^* E_x \\
        0
        \end{array}
    \right)
    \right]  \nonumber \\ &+ \frac{\eta e^3\omega_B^4 \hbar^2  \tau}{960\pi^2\mu^4(1+\omega^2\tau^2)}
    \left[
    \left(
        \begin{array}{c}
        -3(E_y E_z^* + E_y^* E_z) \\
        3(E_x E_z^* + E_x^* E_z) \\
        0
        \end{array}
    \right)+i \omega \tau
    \left(
        \begin{array}{c}
        E_y E_z^* - E_y^* E_z \\
        E_z E_x^* - E_z^* E_x \\
        -2(E_x E_y^* - E_x^* E_y)
        \end{array}
    \right)
    \right]  \nonumber\\ &+ \frac{3 \eta e^3 \omega_B^6 \hbar^3 \tau^2}{160 \pi^2 \mu^5 (1+\omega^2 \tau^2)}  \left(\begin{array}{c} 0 \\ 0 \\ |E_z|^2   \end{array}  \right).
\end{align}
Assuming further the limits $\mu \tau \gg \hbar$ and $\omega \mu^2 \tau^3 \gg \hbar^2$, we can neglect the terms $\propto \omega_B^4/\mu^4$ and reproduce Eqs.~\eqref{semi-mu-limit}-\eqref{Eq:BCDcurrent}. We note that other components, apart from $\sigma^{zzz}_{\text{dc}}$, also receive corrections of order $\mathcal{O}(\omega_B^6)$, however, they are subleading, so we omit them here for brevity.

Analogously, the SHG current in the limit $\omega_B \to 0$ equals
\begin{align}
    \mathbf{j}_2^{2\omega} &\approx \frac{\eta e^3 \omega_B^2 \tau^2 (2-3i \omega\tau )}{24 \pi^2 \mu \hbar (1-i\omega\tau )^2 (1-2i \omega\tau)}
    \left(
        \begin{array}{c}
        E_z E_x \\
        E_z E_y \\
        -(E_x^2 + E_y^2)
        \end{array}
    \right)
    \\ & + \frac{i \eta e^3 \omega_B^4 \omega  \tau ^4}{16 \pi^2 \mu^2 (1-i\omega \tau)}
   \left(\frac{2-3i\omega \tau}{3(1-i\omega \tau)^2 (1-2i\omega\tau)^2} + \frac{i \hbar^2}{20 \omega \mu^2 \tau^3} \right)
    \left(
        \begin{array}{c}
        E_z E_y\\
        -E_z E_x \\
        0
        \end{array}
    \right)  + \frac{3 \eta e^3 \omega_B^6\hbar^3  \tau^2}{320 \pi^2 \mu^5 (1-i\omega \tau)(1-2i\omega\tau)} \left(
        \begin{array}{c}
        0\\
       0 \\
        E_z^2
        \end{array}
    \right). \nonumber
\end{align}
Assuming again $\omega \mu^2 \tau^3 \gg \hbar^2$, we reproduce Eq.~\eqref{semi2w-mu-limit3}. Here we also neglected the contributions of order $\mathcal{O}(\omega_B^6)$ to all components except for $\sigma^{zzz}_{2\omega}$ as subleading. We also note that all the results in this section can be obtained by expanding in powers of $1/\mu$ at $\mu \to \infty$.

\subsection{Clean low-field limit, $\tau\to \infty$ followed by $\omega_B\rightarrow 0$ \label{App:largetausmallB}}

Next, we consider the clean limit $\tau \to \infty$ followed by the low-field limit $\omega_B\to0$. Expanding the exact solution to the order $\mathcal{O}(1/\tau^0)$ and subsequently to the leading order in $\omega_B$, we find for the photocurrent:
\begin{align}
    \mathbf j_2^\text{dc}
    &\approx
    \frac{i \eta e^3}{4\pi^2\omega\hbar^2}
    \left(
        \begin{array}{c}
        E_yE_z^*-E_zE_y^*\\
        E_zE_x^*-E_xE_z^*\\
        \frac23(E_xE_y^*-E_yE_x^*)
        \end{array}
    \right) -\frac{i \eta e^3\omega_B^4
    \left(20\mu^2 + \hbar^2 \omega^2
    \right)}
    {480\pi^2\mu^4\omega^3}
    \left(
        \begin{array}{c}
        0\\
        0\\
        E_xE_y^*-E_yE_x^*
        \end{array}
    \right)
    \nonumber\\
    &
    -\frac{\eta e^3\omega_B^6 \hbar}
    {480\pi^2\mu^5\omega^4}
    \left(
        \begin{array}{c}
        0\\
        0\\
        \left(20 \mu^2 - \hbar^2 \omega^2\right) \left(|E_x|^2 + |E_y|^2\right) - 9 \hbar^2 \omega^2 |E_z|^2
        \end{array}
    \right).
\end{align}
Assuming further $\hbar \omega \ll \mu$ and dropping the terms independent of $\mu$, we reproduce Eq.~\eqref{semi-tauB-limit}.

For the SHG, we obtain: 
\begin{align}
    \mathbf j_2^{2\omega}
    &\approx
    \frac{\eta e^3\omega_B^2}{16\pi^2\mu\omega^2\hbar}
    \left(
        \begin{array}{c}
        -E_xE_z\\
        -E_yE_z\\
        E_x^2+E_y^2
        \end{array}
    \right) +
    \frac{i \eta e^3   \omega_B^4 \left(5 \mu ^2- \hbar^2\omega^2 \right)}
    {320 \pi ^2 \mu ^4 \omega ^3}
    \left(
        \begin{array}{c}
        E_yE_z\\
        -E_xE_z\\
        0
        \end{array}
    \right)
    \nonumber\\
    &
    -
    \frac{\eta e^3   \omega_B^6 \hbar}
    {768 \pi ^2 \mu ^5 \omega ^4}
    \left(
        \begin{array}{c}
       (3\mu^2 - \hbar^2 \omega^2) E_x E_z\\
        (3\mu^2 - \hbar^2 \omega^2) E_y E_z\\
        -12 \mu^2 \left(E_x^2 + E_y^2\right) + \frac{\hbar^2 \omega^2}5 \left[  11\left(E_x^2 + E_y^2 \right) + 18 E_z^2\right]
        \end{array}
    \right),
\end{align}
which in the limit $\hbar \omega \ll \mu$ reproduces Eq.~\eqref{semi2w-tauB-limit} to the leading order in $\omega_B$.

We also note that the same result can be obtained by performing the expansion at $\tau \to \infty$ to the order $\mathcal{O}(1/\tau^0)$
followed by the expansion in powers of $1/\mu$ at $\mu \to \infty$.

\subsection{Clean low-frequency limit, $\tau\to \infty$ followed by $\omega\rightarrow 0$ \label{App:largetausmallw}}
Analogously, we find the asymptotic expressions in the clean limit $\tau \to \infty$ followed by the low-frequency limit $\omega\to0$. Expanding again the exact solution to the order $\mathcal{O}(1/\tau^0)$ and subsequently to the leading order in $\omega$, we obtain for the photocurrent:
\begin{align}
  \mathbf{j}_2^\text{dc}
    &\approx
    \frac{3 \eta e^3 \omega_B^6 \hbar^3}
    {160\pi^2\mu^5\omega^2}
    \left(
        \begin{array}{c}
        0\\
        0\\
        |E_z|^2
        \end{array}
    \right)
    +\frac{i \eta e^3}
    {4\pi^2\omega\hbar^2}
    \left(
        \begin{array}{c}
        E_yE_z^*-E_zE_y^*\\
        E_zE_x^*-E_xE_z^*\\
        0
        \end{array}
    \right)
    -\frac{2\eta e^3\mu}
    {3\pi^2\omega_B^2\hbar^3}
    \left(
        \begin{array}{c}
        0\\
        0\\
        |E_x|^2+|E_y|^2
        \end{array}
    \right)
    \nonumber\\
    &-\frac{2i \eta e^3 \mu^2 \omega}{ \pi ^2 \omega_B^4 \hbar ^4}
    \left(
        \begin{array}{c}
        0\\
        0\\
        E_xE_y^*-E_yE_x^*
        \end{array}
    \right),
\end{align}
where we also kept only the leading-order in $\hbar\omega_B/\mu \ll 1$ terms. Leaving out the terms independent of $\mu$, we reproduce Eq.~\eqref{semi-tauw-limit}.

For the SHG, we find
\begin{align}
    \mathbf j_2^{2\omega}
    &\approx
    -\frac{3\eta e^3 \omega_B^6 \hbar ^3}{640 \pi ^2 \mu ^5 \omega ^2}
    \left(
        \begin{array}{c}
        0\\
        0\\
        E_z^2
        \end{array}
    \right)
    -
    \frac{i \eta e^3}{4\pi^2\omega\hbar^2}
    \left(
        \begin{array}{c}
        E_yE_z\\
        -E_xE_z\\
        0
        \end{array}
    \right)
    +
    \frac{\eta e^3\mu}{\pi^2\omega_B^2\hbar^3}\left(1 - \frac{\hbar^4 \omega_B^4}{80 \mu^4}\right)
    \left(
        \begin{array}{c}
        E_xE_z\\
        E_yE_z\\
        -(E_x^2+E_y^2)/4
        \end{array}
    \right)
    \nonumber\\
    & -\frac{4i\eta e^3 \mu^2 \omega}{\pi^2   \omega_B^4 \hbar ^4}\left(1 - \frac{\hbar^4 \omega_B^4}{80 \mu^4}\right)
    \left(
        \begin{array}{c}
        E_yE_z\\
        -E_xE_z\\
        0
        \end{array}
    \right).
\end{align}
Assuming further $\hbar\omega_B/\mu \ll 1$ and dropping again the terms independent of $\mu$, we reproduce Eq.~\eqref{semi2w-tauomega-limit}.

\subsection{Asymptotic behavior near resonances}

To derive asymptotic expressions for the second-order conductivity components near low-frequency resonances $\omega = \hbar \omega_B^2/(2\mu)$ and $\omega = \hbar \omega_B^2/(4\mu)$, we introduce new dimensionless variables 
\begin{align}
    \lambda \equiv \hbar\omega_B/\mu, \hspace{0.3cm} \gamma \equiv \frac{\mu \omega}{\hbar\omega_B^2}, \hspace{0.3cm} \delta \equiv \omega\tau.
\end{align}
We then expand the exact expressions for the current components~\eqref{jdc} at $\lambda \to 0$, keeping the leading nonvanishing contributions of order $\mathcal{O}(1/\lambda^2)$. The resulting expressions for the dc conductivity components, in terms of the original variables, take the form: 
\begin{align}
    \sigma^{xxz}_\text{dc}&=\sigma^{yyz}_\text{dc} \approx -\frac{2\eta e^3 \mu \omega_B^2 \tau ^2 \left(4 \mu ^2+\tau ^2 \omega_B^4 \hbar ^2\right)}{3 \pi ^2 \hbar \left[16 \mu ^4 \left(1+\omega^2\tau^2\right)^2+8 \mu ^2 \tau ^2 \omega_B^4 \hbar ^2 \left(1-\omega^2\tau^2\right)+\tau ^4 \omega_B^8 \hbar ^4\right]},
    \nonumber\\
    \sigma^{xyz}_\text{dc} &= (\sigma^{yxz}_\text{dc})^* \approx \frac{2 i \eta e^3  \mu ^2 \tau ^2 \omega \left[4 \mu ^2 \left(1+\omega^2\tau^2\right)-3 \tau ^2 \omega_B^4 \hbar ^2\right]}{3 \pi ^2 \hbar ^2 \left[16 \mu ^4 \left(1+\omega^2\tau^2\right)^2+8 \mu ^2 \tau ^2 \omega_B^4 \hbar ^2 \left(1-\omega^2\tau^2\right)+\tau ^4 \omega_B^8 \hbar ^4\right]},
    \nonumber\\
    \sigma^{xzx}_\text{dc}&=\sigma^{yzy}_\text{dc}=(\sigma^{zxx}_\text{dc})^*=(\sigma^{zyy}_\text{dc})^* \approx \frac{\eta e^3 \mu \tau ^2 \omega_B^2 \left[16 \mu ^4 (2 - i \omega \tau - \omega^2 \tau^2)+4 \mu ^2 \tau ^2 \omega_B^4 \hbar ^2 (9 - 3 i \omega \tau - \omega^2 \tau^2)+3\tau ^4 \omega_B^8 \hbar ^4\right]}{6 \pi ^2 \hbar (1+i\omega \tau ) \left(4 \mu ^2+\tau ^2 \omega_B^4 \hbar ^2\right)^2 \left[\tau ^2 \omega_B^4 \hbar ^2+4 \mu ^2 (1-i\omega \tau)^2\right]},
    \nonumber\\
    \sigma^{yzx}_\text{dc}&=-\sigma^{xzy}_\text{dc}=(\sigma^{zyx}_\text{dc})^*=-(\sigma^{zxy}_\text{dc})^*
    \nonumber\\
    &\approx \frac{ \eta e^3 \tau ^2 \left[128 i \mu ^6 \omega (1-i\omega \tau )+16 i \mu ^4 \tau ^2 \omega \omega_B^4 \hbar ^2 (4-5i \omega \tau )+4 \mu ^2 \tau ^3 \omega_B^8 \hbar ^4 (3 \omega^2 \tau^2+ 4 i \omega \tau - 7)-3 \tau ^5 \omega_B^{12} \hbar ^6\right]}{12 \pi ^2 \hbar ^2 (1+i \omega \tau) \left(4 \mu ^2+\tau ^2 \omega_B^4 \hbar ^2\right)^2 \left[\tau ^2 \omega_B^4 \hbar ^2+4 \mu ^2 (1-i\omega \tau )^2\right]}.
\label{finite-gamma-delta}
\end{align}
Assuming further $\omega \tau \gg 1$, we obtain
\begin{align}
    \sigma^{xxz}_\text{dc}&=\sigma^{yyz}_\text{dc} \approx -\frac{2\eta e^3 \mu \omega_B^2 \left(\frac{4 \mu^2}{\tau^2}+\omega_B^4 \hbar ^2\right)}{3 \pi ^2 \hbar \left[\left(4\mu^2 \omega^2 - \omega_B^4 \hbar^2\right)^2 + \frac{8 \mu^2 (4\mu^2 \omega^2 + \omega_B^4 \hbar^2)}{\tau^2} \right]}, \\
    \sigma^{xyz}_\text{dc} &= (\sigma^{yxz}_\text{dc})^* \approx \frac{2 i \eta e^3  \mu ^2 \omega \left[4 \mu ^2 \omega^2-3 \omega_B^4 \hbar^2\right]}{3 \pi^2 \hbar ^2 \left[ \left(4\mu^2 \omega^2 - \omega_B^4 \hbar^2\right)^2 + \frac{8 \mu^2 (4\mu^2 \omega^2 + \omega_B^4 \hbar^2)}{\tau^2}    \right]},
    \nonumber\\
    \sigma^{xzx}_\text{dc}&=\sigma^{yzy}_\text{dc}=(\sigma^{zxx}_\text{dc})^*=(\sigma^{zyy}_\text{dc})^* \approx i\frac{\eta e^3 \mu  \omega_B^2 \left(3 \omega_B^8 \hbar^4 - 4 \mu^2 \omega^2 \omega_B^4 \hbar^2 - \frac{16\mu^4 \omega^2}{\tau^2}\right)}{6 \pi^2 \hbar \omega \tau \left(\frac{4 \mu^2}{\tau^2}+ \omega_B^4 \hbar^2\right)^2 \left(4 \mu^2 \omega^2 - \omega_B^4 \hbar^2 + 8 i \frac{\mu^2 \omega}{\tau}\right)},
    \nonumber\\
    \sigma^{yzx}_\text{dc}&=-\sigma^{xzy}_\text{dc}=(\sigma^{zyx}_\text{dc})^*=-(\sigma^{zxy}_\text{dc})^*
    \approx -\frac{ \eta e^3 \left[3 i \omega_B^8 \hbar^4 \tau \left(\omega_B^4 \hbar^2 - 4\mu^2 \omega^2\right) + 16 \mu^2 \omega \omega_B^8 \hbar^4 - 80i \mu^4 \omega^2 \omega_B^4 \frac{\hbar^2}{\tau} - 128 i \frac{\mu^6 \omega^2}{\tau^3}\right]}{12 \pi^2 \hbar^2 \omega \tau \left(\frac{4 \mu^2}{\tau^2}+ \omega_B^4 \hbar^2\right)^2 \left(4 \mu^2 \omega^2 - \omega_B^4 \hbar^2 + 8 i \frac{\mu^2 \omega}{\tau}\right)}. \nonumber
\end{align}
Focusing on the vicinity of the resonant frequency $\omega = \hbar \omega_B^2/(2\mu)$ and keeping a singular part only, we derive Eq.~\eqref{semi-tau-limit2-v3}. We note that the condition $\omega \tau \gg 1$ near the resonance also implies that $\omega_B^2 \hbar \tau \gg \mu$.

Performing analogous calculations for the SHG components, we find at the order $\mathcal{O}(1/\lambda^2)$
\begin{align}
    \sigma^{xxz}_{2\omega}&=\sigma^{yyz}_{2\omega}\approx -\frac{\eta e^3 \mu \omega_B^2 \tau^2 (2-3i \omega\tau )}{6 \pi ^2 \hbar(1-2i \omega\tau ) \left[\tau ^2 \omega_B^4 \hbar ^2+4 \mu ^2 (1-i\omega\tau )^2\right]},
    \nonumber\\\nonumber
    \sigma^{zxx}_{2\omega} &= \sigma^{xzx}_{2\omega} = \sigma^{zyy}_{2\omega} = \sigma^{yzy}_{2\omega}
    \\
    &\approx \frac{\eta e^3 \mu \tau ^2 \omega_B^2 (1-2i \omega\tau) \left[16 \mu ^4 (1-2i \omega\tau )^2 (1-i\omega \tau)(2-3 i \omega \tau)-4 \mu ^2 \tau ^2 \omega_B^4 \hbar ^2 \left(15 \tau ^2 \omega ^2+23 i \omega\tau -9\right)+3 \tau ^4 \omega_B^8 \hbar ^4\right]}{12 \pi ^2 \hbar (1-i\omega\tau) \left[\tau ^2 \omega_B^4 \hbar ^2+4 \mu ^2 (1-i\omega\tau )^2\right] \left[\tau ^2 \omega_B^4 \hbar ^2+4 \mu ^2 (1-2i \omega\tau)^2\right]^2},
    \nonumber
    \\ \nonumber
    \sigma^{zyx}_{2\omega}&=\sigma^{yzx}_{2\omega}=-\sigma^{zxy}_{2\omega}=-\sigma^{xzy}_{2\omega}
    \\
    &\approx \frac{i \eta e^3  \tau ^4 \omega_B^4 \left[16 \mu ^4 \omega (1-2i \omega \tau)^2 (2-3i \omega \tau )+4 \mu ^2 \omega_B^4 \tau \hbar ^2 \left(-15 i \tau ^2 \omega ^2+20 \omega \tau +7 i\right)+3 i \tau ^3 \omega_B^8 \hbar ^4\right]}{24 \pi ^2 (1-i\omega \tau) \left[\tau ^2 \omega_B^4 \hbar ^2+4 \mu ^2 (1-i\omega \tau)^2\right] \left[\tau ^2 \omega_B^4 \hbar ^2+4 \mu ^2 (1-2i \omega \tau)^2\right]^2}.
\label{finite-gamma-delta_SHG}
\end{align}
Carefully expanding and simplifying these expressions at $\omega \tau \gg 1$, we obtain
\begin{align}
    \sigma^{xxz}_{2\omega}&=\sigma^{yyz}_{2\omega}\approx - \frac{\eta e^3 \mu \omega_B^2}{4\pi^2 \hbar \left(\omega_B^4 \hbar^2 - 4\mu^2 \omega^2 - 8i\frac{\mu^2 \omega}{\tau}  \right)},
    \nonumber\\\nonumber
    \sigma^{zxx}_{2\omega} &= \sigma^{xzx}_{2\omega} = \sigma^{zyy}_{2\omega} = \sigma^{yzy}_{2\omega}
    \\
    &\approx \frac{\eta e^3 \mu \omega_B^2}{2\pi^2 \hbar \left(\omega_B^4 \hbar^2 - 16\mu^2 \omega^2 - 16i\frac{\mu^2 \omega}{\tau}  \right)} - \frac{2 i \eta e^3 \mu^3 \omega_B^2 \omega \left(16 \omega^2 \mu^2 + 5 \hbar^2 \omega_B^4 \right)}{3 \pi^2 \hbar \tau \left(\omega_B^4 \hbar^2 - 16\mu^2 \omega^2 - 16i\frac{\mu^2 \omega}{\tau}  \right)^2 \left(\omega_B^4 \hbar^2 - 4\mu^2 \omega^2 - 8i\frac{\mu^2 \omega}{\tau}  \right) },
    \nonumber
    \\ \nonumber
    \sigma^{zyx}_{2\omega}&=\sigma^{yzx}_{2\omega}=-\sigma^{zxy}_{2\omega}=-\sigma^{xzy}_{2\omega}
    \\
    & \approx -\frac{i\eta e^3 \omega_B^4}{8\pi^2 \omega\left(\omega_B^4 \hbar^2 - 16\mu^2 \omega^2 - 16i\frac{\mu^2 \omega}{\tau}  \right)} - \frac{ \eta e^3 \mu^2 \omega_B^4 \left(32 \omega^2 \mu^2 + \hbar^2 \omega_B^4 \right)}{3 \pi^2 \tau \left(\omega_B^4 \hbar^2 - 16\mu^2 \omega^2 - 16i\frac{\mu^2 \omega}{\tau}  \right)^2 \left(\omega_B^4 \hbar^2 - 4\mu^2 \omega^2 - 8i\frac{\mu^2 \omega}{\tau}  \right) }.
\end{align}
Focusing on the vicinity of the resonance frequency $\omega=\hbar\omega_B^2/2\mu$ for the components $\sigma^{xxz}_{2\omega}=\sigma^{yyz}_{2\omega}$, and $\omega=\hbar\omega_B^2/4\mu$ for all remaining components, we derive Eq.~\eqref{semi2w-tau-limit2}. We note that the latter components do not exhibit a singularity or peak at $\omega=\hbar\omega_B^2/2\mu$.

Finally, we note in passing that the results of Appendices~\ref{App:largetausmallB} and~\ref{App:largetausmallw} under the additional assumptions $\hbar \omega \ll \mu$ and $\hbar \omega_B \ll \mu$, correspondingly, can be readily derived from Eqs.~\eqref{finite-gamma-delta} and~\eqref{finite-gamma-delta_SHG}. Indeed, rewriting these equations in variables $\{\lambda,\gamma,\delta\}$ and expanding them at $\delta \to \infty$ to order $\mathcal{O}(1/\delta^0)$ and then at $\gamma \to \infty$ to order $\mathcal{O}(1/\gamma^4)$, we reproduce the clean low-field limit described by Eqs.~\eqref{semi-tauB-limit} and~\eqref{semi2w-tauB-limit}, except for the $zzz-$components. Analogously, expanding Eqs.~\eqref{finite-gamma-delta} and~\eqref{finite-gamma-delta_SHG} at $\delta \to \infty$ to order $\mathcal{O}(1/\delta^0)$ and then at $\gamma \to 0$ to order $\mathcal{O}(\gamma)$, we reproduce the clean low-frequency limit described by Eqs.~\eqref{semi-tauw-limit} and~\eqref{semi2w-tauomega-limit}, except for the $zzz-$components. To reproduce the $zzz$-components, one should probably expand the exact expressions for the current to higher orders in $\lambda$, although we have not performed this calculation. 

\end{widetext}

\bibliography{ShiftCurrent}

\end{document}